\documentclass[useAMS,usenatbib,onecolumn]{mn2e}
\usepackage{epsfig}
\usepackage{amssymb,latexsym,amsopn}
\usepackage{natbib}
\usepackage{graphicx}
\usepackage[T1]{fontenc}
\usepackage{ulem}
\title[Synchrotron emission from pulsars]
{Synchrotron X-ray emission from old pulsars}
\author[Kisaka \& Tanaka]
{Shota Kisaka$^{1}$\thanks{kisaka@post.kek.jp} and Shuta J. Tanaka$^{2}$\thanks{sjtanaka@icrr.u-tokyo.ac.jp}\\
$^{1}$ Institute of Particle and Nuclear Studies, KEK, Tsukuba, Ibaraki, 305-0801, Japan \\ 
$^{2}$ Institute for Cosmic Ray Research, University of Tokyo, Kashiwa, Chiba, 277-8582, Japan \\
}

\date{\today}
\begin{document}

\maketitle
\begin{abstract}

We study the synchrotron radiation as the observed non-thermal X-ray emission from old pulsars ($\gtrsim1-10$Myr) to investigate the particle acceleration in their magnetospheres. We assume that the power-law component of the observed X-ray spectra is caused by the synchrotron radiation from electrons and positrons in the magnetosphere. We consider two pair production mechanisms of X-ray emitting particles, the magnetic and the photon-photon pair productions. High-energy photons, which ignite the pair production, are emitted via the curvature radiation of the accelerated particles. We use the analytical description for the radiative transfer and estimate the luminosity of the synchrotron radiation. We find that for pulsars with the spin-down luminosity $L_{\rm sd}\lesssim10^{33}$ erg s$^{-1}$, the locations of the particle acceleration and the non-thermal X-ray emission are within $\lesssim10^7$cm from the centre of the neutron star, where the magnetic pair production occurs. For pulsars with the spin-down luminosity $L_{\rm sd}\lesssim10^{31}$ erg s$^{-1}$ such as J0108-1431, the synchrotron radiation is difficult to explain the observed non-thermal component even if we consider the existence of the strong and small-scale surface magnetic field structures. 

\end{abstract}
%
\section{INTRODUCTION}

Pulsars emit high-energy photons powered by their rotational energy loss. It is considered that particles are significantly accelerated and emit $\gamma$-ray in a gap, where the charge density is depleted from the Goldreich-Julian charge density \citep{GJ69}. The phase-averaged spectrum above 200 MeV observed by {\it Fermi} rules out the near-surface acceleration proposed in polar cap cascade models \citep{Ab09}, which would exhibit a much sharp spectral cutoff due to the magnetic pair-production attenuation \citep{DH96}. Thus, accelerated particles and their $\gamma$-ray emission originate in the outer magnetosphere, as considered in the outer gap model \citep{CHR86}.

In the outer gap model, the gap is self-sustained by the development of the electromagnetic cascade near the null charge surface. \citet{WH11} suggested a certain limit for the combination of rotational period and its time derivative, which the outer gap becomes no longer self-sustained (in their figure 2). As a pulsar gets old and spins down, the null charge surface in the open magnetosphere gets further away from the star so that the efficiency of the particle production decreases in the outer gap model. Old pulsars which we define as non-recycled pulsars with characteristic ages $\gtrsim1-$10Myr in this paper are beyond the limit suggested by \citet{WH11}, and have not been detected in $\gamma$-ray yet \citep{Ab13}.

A dozen of old pulsars have been detected in X-ray band (e.g., Posselt et al. 2012b). Although the photon statistics is limited, the observed X-ray spectra can be fitted by a combination of thermal and non-thermal components (e.g., Becker et al. 2004). The observed flux ratio of non-thermal to thermal components in X-ray band is an order of or larger than unity for most old pulsars. Thermal X-ray component is generally considered as a combination of the emission from the bulk surface of the neutron star and the polar cap heated by the infall of accelerated and produced particles \citep{HR93, HM01, HM02}. For the old neutron stars, the surface temperature is too cold to detect the significant thermal X-ray emission from the bulk surface of the neutron star \citep{YP04}. In fact, using a simple blackbody model to fit the thermal component, obtained temperatures, ($\sim$1-3) $\times10^6$ K, is consistent with the prediction of the heated polar cap (e.g., Halpern et al. 1993), and the projected areas are much smaller than the bulk neutron star surface (e.g., Zavlin \& Pavlov 2004). Therefore, the detected thermal component is only considered as the heated polar cap for old pulsars. The non-thermal X-ray efficiencies, which are the ratio between the observed X-ray luminosity in the 1-10keV band to the spin-down luminosity are about $\sim10^{-2}$-$10^{-5}$ \citep{Ka12}. Non-thermal component is considered as the radiation from particles produced during the process of the electromagnetic cascade in the gap (e.g., Cheng et al. 1986; Zhang \& Harding 2000). Based on the scenario for both thermal and non-thermal components, the significant particle acceleration should occur in somewhere of the old pulsar magnetospheres.

For the non-thermal X-ray emission mechanism from young pulsars ($\lesssim1$Myr), the synchrotron radiation from second and higher generation particles is considered (e.g., Cheng et al. 1986). The synchrotron radiation models require the occurrence of the significant particle acceleration in the outer magnetosphere (e.g., Zhang \& Cheng 1997; Cheng, Gil \& Zhang 1998). It explains the observed energy spectra and light curves of the non-thermal component (e.g., Takata, Chang \& Shibata 2008; Kisaka \& Kojima 2011). However, as mentioned above, it is suggested that the significant number of particles are not accelerated at the outer magnetosphere for old pulsars \citep{WH11}. Some authors have considered that the particle acceleration region locates at the magnetic pole (polar cap model) for old pulsars \citep{CZ99, WH11}. In the polar cap model, it is predicted that there is a low-energy turnover at the local cyclotron frequency in the X-ray spectra of pulsars \citep{OS70, HD99, RD99}. This indicates that the flux of the synchrotron radiation is significantly decreased in X-ray band. \citet{ZH00} suggested that in the polar cap model, the origin of the non-thermal X-ray emission are the inverse Compton scattering of the higher generation pairs. They found that the efficiency of the converting particle kinetic energy to the radiation by the inverse Compton scattering would be very high if the scatterings occur in the resonant regime. Their model explains the observed luminosity of two old pulsars, B0950+08 and B1929+10. The polar cap model has possibilities to reproduce the observed non-thermal X-ray from old pulsars. Therefore, in order to reproduce the observed X-ray emission, it is expected that as pulsars get old, the significant particle acceleration occurs at inner magnetospheres. 

It is highly unknown how to evolve the location of particle acceleration and non-thermal X-ray emission regions. The synchrotron radiation and the inverse Compton scattering are considered for the mechanisms of the non-thermal X-ray emission. In the case of the synchrotron radiation, because of the high cyclotron frequency at the surface of the neutron star, the emission region of the synchrotron radiation needs to be relatively far from the surface. Newly produced pairs obtain the non-zero pitch angle and emit synchrotron radiation. Then, the synchrotron radiation region corresponds to the particle production region. To supply the large number of $\gamma$-ray photons at the outer region, the particle acceleration region also locates relatively far from the neutron star surface. The number of produced pairs decreases at the outer region because the magnitude of the magnetic field is weak and the number of thermal photons from the neutron star decreases. In order to reproduce the observed luminosity, there are some limits for the location of the synchrotron radiation region. In principle, the particle acceleration region is able to locate between the neutron star surface and the null charge surface. For example, the position of its inner boundary depends on the current density and does not always correspond to the null charge surface \citep{Ta06, H06}. Then, the inner boundary may move close to the neutron star surface for some values of the current density. On the other hand, in the case of the inverse Compton scattering, the particle acceleration region locate near the surface of the neutron star for the model of \citet{ZH00}. Considering models based on two emission mechanisms, the upper limit of the location of the particle acceleration should be determined by the condition whether the synchrotron radiation from the higher generation particles explains the observed non-thermal X-ray emission or not. Therefore, it is worthwhile to determine the limit (death line) that the synchrotron radiation explains the observed non-thermal X-ray emission. Since old pulsars are detected in X-ray but not in $\gamma$-ray, we may be able to obtain the signature how to evolve the location of the particle acceleration from the X-ray observations. 

In this paper, we study the non-thermal X-ray emission from old pulsars. We focus on the synchrotron radiation caused by the second and higher generation particles in the magnetosphere. We consider two pair production processes, the magnetic and the photon-photon ones, separately. We also consider two directions of the primary particles, outward (away from the neutron star) and inward (toward to the neutron star), so that four models are investigated. We parameterize the distances from the centre of the neutron star to the synchrotron and the curvature radiation regions and give some constraints for them. In section 2, we introduce our models and several constraints for the location of the emission region. In sections 3 and 4, we present the death lines to reproduce the observed non-thermal X-ray emission for four type of models. Discussion is presented in section 5. Conclusions are given in section 6.

\section{MODEL}

\subsection{Assumptions}

We assume that X-ray spectrum from old pulsars consists of both thermal and non-thermal components. We adopt the scenario that the thermal luminosity corresponds to the kinetic energy flux of the ingoing primary particles (e.g., Halpern \& Ruderman 1993). This scenario is supported by the observed temperature and the emitting area \citep{YP04, ZP04}. For the non-thermal component, the synchrotron radiation from secondary particles is considered.

In our definitions, "primary particles" mean the electrons and positrons that are accelerated by the magnetic-field-aligned electric field and emit curvature photons that can convert pairs. "Secondary particles" mean those produced outside the acceleration region, including second and higher generation particles. The production and emission locations for second and higher generation particles are almost the same in our model. We focus on non-thermal X-ray emitting particles, regardless the number of generation. In these reasons, we do not separately treat second and higher generation particles. 

Throughout the paper, we fix the mass and radius of neutron star as $M_{\rm NS}=1.4M_{\odot}$ and $R_{\rm NS}=10^6$cm. We only consider that the emission region of the synchrotron radiation locates within the radius of the light cylinder. For the global magnetic structure of the old pulsar magnetospheres, we assume the dipole field. We briefly discuss the effects of the non-dipole magnetic field in section 5.1. Note that we mainly consider the inner region of the magnetosphere, so that our model does not significantly depend on the inclination angle between the rotation and the magnetic axes. Since the location of the particle acceleration is highly uncertain for the old pulsar magnetospheres, we parameterize the distance from the centre of the neutron star to the emission location of the curvature radiation from primary particles, $r_{\rm pri}$. We also parameterize the distance from the centre of the neutron star to the emission region of the synchrotron radiation from secondary particles as $r_{\rm sec}$. In our model, the locations of the particle production and the synchrotron radiation are the same location due to the smaller timescale of the synchrotron radiative cooling than that of the advection in most regions of the magnetosphere. We constrain $r_{\rm pri}$ and $r_{\rm sec}$ from the observational results.

In order to compare our results with the observations, we select seven old pulsars whose X-ray spectrum were analyzed and the values of the observed thermal luminosity $L_{\rm th}$, non-thermal luminosity $L_{\rm syn}$ and temperature of the heated polar cap $T_{\rm pc}$ were published in the literatures. We summarize these parameters in table 1. We evaluate the magnitude of the magnetic field at the neutron star surface, $B_{\rm s}$, which we use the relation of the magnetic moment of the neutron star assuming the vacuum dipole radiation formula, $\mu_{\rm mag}^2=3Ic^3P\dot{P}/8\pi^2$, where $I$ is the moment of inertia. We introduce a fiducial old pulsar with $P_0=B_{\rm s,12}=L_{\rm sd,31}=\nu_{\rm obs,keV}=L_{\rm syn,28}=L_{\rm th,28}=T_{\rm pc,6}=1$, where $P$ is the rotational period, $L_{\rm sd}$ is the spin-down luminosity and $\nu_{\rm obs}$ is the observed frequency. Hereafter, we use $Q_x\equiv Q/10^x$ in cgs units, and $h\nu_{\rm keV}\equiv h\nu/1$keV. 

We consider two particle production mechanisms for the secondary particles. In the outer magnetosphere, secondary particles are produced by the photon-photon pair production so that we need to consider this process. In addition, we also need to consider the magnetic pair production for old pulsars because of following reasons. There is the possibility that the inwardly accelerated particles attain near the star without the significant deceleration. Then, the curvature photons emitted from them are materialized by magnetic pair production \citep{ZC97}. There is an other possibility that as a pulsar gets old, the gap region may move toward the neutron star surface to produce the enough number of particles to sustain the gap. Once the inner boundary of the gap moves toward the neutron star surface enough to occur the magnetic pair production, this process should dominate the particle production. We treat two production mechanisms separately. 

For the moving direction of primary particles, we consider both outward (section 3) and inward directions (section 4), independently. The maximum value of the kinetic energy flux of primary particles and the efficiency of the particle production are different between the outward and the inward cases. The maximum value of the kinetic energy flux corresponds to the spin-down luminosity $L_{\rm sd}$ for the outward case and to the observed thermal luminosity from the heated polar cap $L_{\rm th}$ for the inward case. We consider that the kinetic energy fluxes between outward and inward directions can take different values (Timokhin 2010). We take these values for each case because we demand the maximum radiative efficiency of the synchrotron radiation for old pulsars to obtain the limitation for the locations of the particle acceleration and the X-ray emission. The particle production efficiency is also different between the cases of the outgoing and the ingoing primary particles. For the magnetic pair production, the efficiency strongly depends on the angle $\theta_{\rm B\gamma}$ between the propagation direction of the curvature photons emitted by a primary particle and the direction of the local magnetic field \citep{E66}. The efficiency for the photon-photon pair production also depends on the collision angle $\theta_{\rm col}$ between the curvature photon and the thermal X-ray photon from the heated polar cap. For the outgoing case, the angles $\theta_{\rm B\gamma}$ and $\theta_{\rm col}$ are expected to be very small near the neutron star surface (section 3). On the other hand, for the ingoing case, the angles $\theta_{\rm B\gamma}$ and $\theta_{\rm col}$ are expected to be large value in principle (section 4). 

\subsection{Constraints}

\subsubsection{Energy of secondary particles}

We assume that the secondary particles are produced by the curvature photons emitted by the primary particles. We obtain the Lorentz factor $\gamma_{\rm s,pair}(r_{\rm pri})$ of produced pairs as following. In this section, in order to clarify which value of distance ($r_{\rm sec}$ or $r_{\rm pri}$) we use for the calculations, we obviously denote arguments to physical quantities which depend on them. 

The characteristic energy of the curvature radiation is determined by the Lorentz factor $\gamma_{\rm p}$ of the primary particles and the curvature radius $R_{\rm cur}(r_{\rm pri})$ of the field line at the particle acceleration region,
\begin{eqnarray}\label{sec2:E_cur}
E_{\rm cur}(r_{\rm pri})=0.29\frac{3h\gamma_{\rm p}^3c}{4\pi R_{\rm cur}(r_{\rm pri})},
\end{eqnarray}
where $c$ is speed of light and $h$ is Planck constant. From the assumption of the dipole magnetic field, we use the approximation $R_{\rm cur}(r_{\rm pri})\sim(r_{\rm pri}R_{\rm lc})^{1/2}$ as a curvature radius of a field line, where $R_{\rm lc}=Pc/2\pi\sim4.8\times10^9P_0$ cm is the radius of the light cylinder. The value of $\gamma_{\rm p}$ should be determined by the force balance between the radiation reaction force and the electric field acceleration (e.g., Cheng et al. 1986). The accelerating electric field is the parallel component to the magnetic field. To calculate $\gamma_{\rm p}$ from the force balance, the distribution of the accelerating electric field in the magnetosphere is required. However, the electric field structure of the old pulsar magnetosphere is highly uncertain. In this paper, we use the maximum value of Lorentz factor $\gamma_{\rm p}$ using the potential drop across the polar cap, $B_{\rm s}(2\pi/P)^2R_{\rm NS}^3/2c^2$, as
\begin{eqnarray}\label{sec2:gamma_p}
\gamma_{\rm p}&=&\frac{2\pi^2eB_{\rm s}R_{\rm NS}^3}{m_{\rm e}c^4P^2},
\end{eqnarray}
where $e$ and $m_{\rm e}$ are the charge and mass of an electron. This assumption gives the maximum energy of secondary particles. We evaluate the Lorentz factor $\gamma_{\rm s,pair}(r_{\rm pri})$ of secondary particles as
\begin{eqnarray}\label{sec2:gamma_s}
\gamma_{\rm s,pair}(r_{\rm pri})=\frac{E_{\rm cur}(r_{\rm pri})}{2m_{\rm e}c^2}.
\end{eqnarray}
This is a theoretical estimate of the energy of secondary particles adopted in our model. Since the energy $E_{\rm cur}(r_{\rm pri})$ is the decreasing function with the distance $r_{\rm pri}$, the value of Lorentz factor $\gamma_{\rm s,pair}(r_{\rm pri})$ is also the decreasing function with the distance $r_{\rm pri}$.

On the other hand, we estimate the observationally required Lorentz factor $\gamma_{\rm s,syn}(r_{\rm sec})$ of secondary particles using the observed frequency $\nu_{\rm obs}$. The characteristic frequency of the synchrotron radiation $\nu_{\rm syn}(r_{\rm sec})$ relates with the magnitude of the magnetic field $B(r_{\rm sec})$ at the emission region of the synchrotron radiation from the secondary particles and the Lorentz factor of the secondary particles $\gamma_{\rm s,syn}(r_{\rm sec})$ as $\nu_{\rm syn}(r_{\rm sec})=0.29(3/4\pi)\gamma_{\rm syn}^2(r_{\rm sec})eB(r_{\rm sec})\alpha(r_{\rm sec})/(m_{\rm e}c)$, where we approximate the pitch angle $\sin\alpha(r_{\rm sec})\sim\alpha(r_{\rm sec})~(<1)$. Using this description and $\nu_{\rm syn}(r_{\rm sec})\rightarrow\nu_{\rm obs}$, we estimate the value of the Lorentz factor of the secondary particles as
\begin{eqnarray}\label{sec2:nu_syn}
\gamma_{\rm s,syn}(r_{\rm sec})=\sqrt{\frac{4\pi}{0.87}\nu_{\rm obs}\frac{m_{\rm e}c}{eB(r_{\rm sec})\alpha(r_{\rm sec})}}.
\end{eqnarray}
The synchrotron approximation breaks down for $\gamma_{\rm s,syn}(r_{\rm sec})\alpha(r_{\rm sec})<1$. The condition $\gamma_{\rm s,syn}(r_{\rm sec})\alpha(r_{\rm sec})\sim1$ gives a cyclotron turnover frequency
\begin{eqnarray}\label{sec2:nu_ct}
\nu_{\rm ct}(r_{\rm sec})=\frac{eB(r_{\rm sec})}{2\pi m_{\rm e}c\alpha(r_{\rm sec})}.
\end{eqnarray}
This frequency is the lower limit for the characteristic frequency of the conventional synchrotron radiation. 

Since the small pitch angle ($\sim10^{-2}$) and the strong magnetic field ($\sim10^{12}$G) are expected in the pulsar magnetosphere, the value of $h\nu_{\rm ct}(r_{\rm sec})$ reaches $\sim1$MeV at the star surface \citep{OS70, RD99}. Then, the perpendicular momentum of the X-ray emitting particles due to the synchrotron radiation is the non-relativistic regime $\gamma_{\rm s,syn}(r_{\rm sec})\alpha(r_{\rm sec}) m_{\rm e}c\lesssim m_{\rm e}c$ near the neutron star surface. If perpendicular momentum is initially relativistic value, subsequent small-pitch-angle synchrotron radiation (after losing perpendicular momentum) from this particle will be negligible compared to the conventional synchrotron radiation for fixed energy band (Epstein 1973). However, we also consider the higher generation particles as non-thermal X-ray emitting particles which are possible to have initially non-relativistic perpendicular momentum (e.g., Rudak \& Dyks 1999). In this case, it is valuable to investigate the contribution of X-ray luminosity due to small-pitch-angle synchrotron radiation. In \citet{E73} the synchrotron radiation by particles with very small pitch angles ($\gamma\alpha\ll1$) is discussed in detail. The characteristic frequency for the small-pitch-angle synchrotron radiation $\nu_{\rm spa}(r_{\rm sec})$ is described as $\nu_{\rm spa}(r_{\rm sec})=2\gamma_{\rm s,spa}(r_{\rm sec})eB(r_{\sec})/(2\pi m_{\rm e}c)$. Using this description and $\nu_{\rm spa}(r_{\rm sec})\rightarrow\nu_{\rm obs}$, we estimate the Lorentz factor $\gamma_{\rm s,spa}(r_{\rm sec})$ as
\begin{eqnarray}\label{sec2:nu_spa} 
\gamma_{\rm s,spa}(r_{\rm sec})=\nu_{\rm obs}\frac{\pi m_{\rm e}c}{eB(r_{\rm sec})}.
\end{eqnarray}

In order to reproduce the observed frequency of the non-thermal emission for the synchrotron radiation, the following relation must be satisfied,
\begin{eqnarray}\label{sec2:energy}
\gamma_{\rm s,pair}(r_{\rm pri})>\left\{ \begin{array}{ll}
\gamma_{\rm s,syn}(r_{\rm sec}) & [\nu_{\rm obs} > \nu_{\rm ct}(r_{\rm sec})] \\
\gamma_{\rm s,spa}(r_{\rm sec}) & [\nu_{\rm obs} < \nu_{\rm ct}(r_{\rm sec})]. \\
\end{array} \right.
\end{eqnarray}
Since the magnitude of the magnetic field is the decreasing function with the distance $r_{\rm sec}$, the values of the Lorentz factor $\gamma_{\rm s,syn}(r_{\rm sec})$ and $\gamma_{\rm s,spa}(r_{\rm sec})$ are the increasing function with the distance $r_{\rm sec}$. On the other hand, since the value of the Lorentz factor $\gamma_{\rm s,pair}(r_{\rm pri})$ is the decreasing function with the distance $r_{\rm pri}$ and the values of the Lorentz factor $\gamma_{\rm s,syn}(r_{\rm sec})$ and $\gamma_{\rm s,spa}(r_{\rm sec})$ are the increasing function with the distance $r_{\rm sec}$, i.e., inequality (\ref{sec2:energy}) gives the upper limit for the emission region of the synchrotron radiation. We will discuss each mechanism of the synchrotron radiation separately when we consider the small value of the pitch angle $\alpha(r_{\rm sec})\ll1$.

Note that the emission spectrum of the synchrotron radiation extends down to the lower frequency than the characteristic frequency, $\nu_{\rm ct}(r_{\rm sec})$. Photon index at the frequency range $\nu<\nu_{\rm ct}(r_{\rm sec})$ is consistent with that of low-energy tail of the conventional synchrotron radiation emitted by a single electron when all particles satisfy the condition $\gamma\alpha>1$. This photon index is much harder than that of all X-ray-detected pulsars \citep{B09}, so that any energy distribution of particles with $\gamma\alpha>1$ does not reproduce the observed photon index at the frequency range $\nu_{\rm obs}<\nu_{\rm ct}(r_{\rm sec})$. Therefore, we only consider the small-pitch-angle synchrotron radiation at the frequency range $\nu_{\rm obs}<\nu_{\rm ct}(r_{\rm sec})$.

\subsubsection{Characteristic frequency}

As memtioned above, equation (\ref{sec2:nu_syn}) and the condition $\gamma\alpha>1$ gives a lower limit for the characteristic frequency in the conventional synchrotron radiation, 
\begin{eqnarray}\label{sec2:frequency}
\nu_{\rm obs} \gtrsim \nu_{\rm ct}.
\end{eqnarray}

On the other hand, equation (\ref{sec2:nu_spa}) and the condition $\gamma>1$ indicate that there are a lower limit for the characteristic frequency in the small-pitch-angle synchrotron model. This is given as
\begin{eqnarray}\label{sec2:frequency2}
\nu_{\rm obs} \gtrsim \frac{eB(r_{\rm sec})}{2\pi m_{\rm e}c}.
\end{eqnarray}
The derived limit for the emission region $r_{\rm sec}$ is determined only with the observed frequency $\nu_{\rm obs}$ and the surface magnetic field $B_{\rm s}$.

\subsubsection{Pair production threshold}

We consider two pair production processes, the magnetic and the photon-photon pair productions. For the magnetic pair production, the mean free path $l_{\rm p}(r_{\rm pri},r_{\rm sec})$ strongly depends on the energy of the curvature photons $E_{\rm cur}(r_{\rm pri})$, the magnitude of the magnetic field at the production point of the secondary particles $B(r_{\rm sec})$ and the angle $\theta_{\rm B\gamma}(r_{\rm sec})$ between the directions of the propagation for a curvature photon and the magnetic field at the point $r_{\rm sec}$. We take advantage of this strong dependence to derive the production threshold following the seminal work of \citet{RS75}. The mean free path $l_{\rm p}(r_{\rm pri},r_{\rm sec})$ of a photon of energy $E_{\rm cur}(r_{\rm pri})>2m_{\rm e}c^2$ moving through a region of the magnetic field is \citep{E66, RS75}
\begin{eqnarray}\label{sec2:mfp}
l_{\rm p}(r_{\rm pri},r_{\rm sec})&=&4.4\frac{\hbar c}{e^2}\frac{\hbar}{m_{\rm e}c}\frac{B_{\rm q}}{B_{\perp}(r_{\rm sec})}\exp\left(\frac{4}{3\chi(r_{\rm pri}, r_{\rm sec})}\right)\ \ [\chi(r_{\rm pri},r_{\rm sec})\ll1], \nonumber \\
\chi(r_{\rm pri}, r_{\rm sec})&\equiv&\frac{E_{\rm cur}(r_{\rm pri})}{2m_{\rm e}c^2}\frac{B_{\perp}(r_{\rm sec})}{B_{\rm q}}.
\end {eqnarray}
Here the magnetic fields $B_{\rm q}=m_{\rm e}^2c^3/e\hbar\sim4.4\times10^{13}{\rm G}$ and $B_{\perp}(r_{\rm sec})=B(r_{\rm sec})\sin\theta_{\rm B\gamma}(r_{\rm sec})$. Since the angle $\theta_{\rm B\gamma}(r_{\rm sec})$ reflects to the pitch angle of produced particles, we take $\theta_{\rm B\gamma}(r_{\rm sec})\sim\alpha(r_{\rm sec})$. From equation (\ref{sec2:mfp}), the small decrease in $\chi(r_{\rm pri},r_{\rm sec})$ corresponds to the exponentially large increase in $l_{\rm p}(r_{\rm pri},r_{\rm sec})$ \citep{RS75}. Following \citet{RS75}, we take the critical value for the pair production threshold as
\begin{eqnarray}\label{sec2:Bcreation}
\chi(r_{\rm pri}, r_{\rm sec})>1/15.
\end{eqnarray}
Inequality (\ref{sec2:Bcreation}) gives the upper limit for the emission region in the case of the magnetic pair production.

For the photon-photon pair production process, we consider the thermal photons of the energy $E_{\rm X}$ from the heated polar cap region as the seed photons. The threshold energy for the photon-photon pair production is given by
\begin{eqnarray}\label{sec2:gcreation}
[1-\cos\theta_{\rm col}(r_{\rm sec})]E_{\rm X}E_{\rm cur}(r_{\rm pri})>(m_{\rm e}c^2)^2,
\end{eqnarray}
where $\theta_{\rm col}(r_{\rm sec})$ is the angle between the propagating directions of a curvature photon and a thermal photon. For the energy of the seed photons, we use the observed temperature $T_{\rm pc}$ of the thermal component as $E_{\rm X}\sim2.8kT_{\rm pc}$, where $k$ is Boltzmann constant. The dependence of the angle $\theta_{\rm col}(r_{\rm sec})$ on $r_{\rm sec}$ determines whether inequality (\ref{sec2:gcreation}) gives the upper limit or the lower limit for the distances $r_{\rm pri}$ and $r_{\rm sec}$.

\subsubsection{Non-thermal luminosity}

We require that the number of the produced secondary particles which emit the synchrotron radiation is large enough to reproduce the observed luminosity of the non-thermal component $L_{\rm syn}$. In our model, we estimate the non-thermal luminosity from the products of the power of the synchrotron radiation from a single particle $P_{\rm syn}(r_{\rm sec})$ and the number of the secondary particles contributing the luminosity. The power of the synchrotron radiation $P_{\rm syn}(r_{\rm sec})$ is described as 
\begin{eqnarray}\label{sec2:N_s}
P_{\rm syn}(r_{\rm sec})={\displaystyle\frac{2e^4B^2(r_{\rm sec})\alpha^2(r_{\rm sec})}{3c^3m_{\rm e}^2}}\times\left\{ \begin{array}{ll}
\gamma_{\rm s,syn}^2(r_{\rm sec}) & [\nu_{\rm obs} > \nu_{\rm ct}(r_{\rm sec})] \\
 & \\
\gamma_{\rm s,spa}^2(r_{\rm sec}) & [\nu_{\rm obs} < \nu_{\rm ct}(r_{\rm sec})]. \\
\end{array} \right.
\end{eqnarray}
The effective number of the secondary particles is obtained from the products of the optical depth for the curvature photons $\tau(r_{\rm sec})~(<1)$, the number of the curvature photons emitted by a electron (or a positron) $N_{\gamma}(r_{\rm pri},r_{\rm sec})$ and the effective number of the primary particles $N_{\rm p}(r_{\rm sec})$. 

Since the magnetic pair production is very efficient once it is triggered, we assume the optical depth of the magnetic pair production,
\begin{eqnarray}\label{sec2:tau_Bgamma}
 \tau_{\rm B\gamma}\sim1,
\end{eqnarray} 
as long as the pair production condition, inequality (\ref{sec2:Bcreation}), is satisfied. On the other hand, the pair conversion is not so efficient for the photon-photon pair production process. We evaluate its optical depth $\tau_{\gamma\gamma}(r_{\rm sec})$ as 
\begin{eqnarray}\label{sec2:tau_gammagamma}
\tau_{\gamma\gamma}(r_{\rm sec})&\sim&\frac{L_{\rm th}}{4\pi r_{\rm sec}^2cE_{\rm X}}\sigma_{\gamma\gamma}[1-\cos\theta_{\rm col}(r_{\rm sec})] r_{\rm sec}
\end{eqnarray}
where we use the cross section for the photon-photon pair production $\sigma_{\gamma\gamma}\sim0.2\sigma_{\rm T}$ and $\sigma_{\rm T}$ is Thomson cross section. The luminosity $L_{\rm th}$ is that of the observed thermal component.  

We evaluate the effective number of the curvature photons $N_{\gamma}(r_{\rm pri},r_{\rm sec})$ emitted by a single electron as
\begin{eqnarray}\label{sec2:N_gamma}
N_{\gamma}(r_{\rm pri}, r_{\rm sec})\sim \dot{N}_{\gamma}(r_{\rm pri})t_{\rm ad}(r_{\rm sec})\sim\frac{P_{\rm cur}(r_{\rm pri})}{E_{\rm cur}(r_{\rm pri})}t_{\rm ad}(r_{\rm sec}),
\end{eqnarray}
where $P_{\rm cur}(r_{\rm pri})$ is the emitted power by a single electron,
\begin{eqnarray}\label{sec2:P_cur}
P_{\rm cur}(r_{\rm pri})=\frac{2e^2c}{3R_{\rm cur}^2(r_{\rm pri})}\gamma_{\rm p}^4.
\end{eqnarray}
In the derivation of equation (\ref{sec2:N_gamma}), we assume that the primary particles continuously emit the curvature radiation during the advection timescale of the secondary particles,
\begin{eqnarray}\label{sec2:t_ad}
t_{\rm ad}(r_{\rm sec})\sim \frac{r_{\rm sec}}{c}.
\end{eqnarray}
From equations (\ref{sec2:E_cur}) and (\ref{sec2:P_cur}), the number $N_{\gamma}(r_{\rm pri},r_{\rm sec})$ is proportional to the Lorentz factor $\gamma_{\rm p}$. 

In order to estimate the effective number of the primary particles which move with the outward direction from the star, we assume that the kinetic energy flux of the primary particles $\dot{N}_{\rm p,out}\gamma_{\rm p}m_{\rm e}c^2$ equal to the spin-down luminosity $L_{\rm sd}$,
\begin{eqnarray}\label{sec2:dotN_p,out}
\dot{N}_{\rm p,out}=\frac{L_{\rm sd}}{\gamma_{\rm p}m_{\rm e}c^2}.
\end{eqnarray}
The spin-down luminosity corresponds to the maximum kinetic energy flux for the outgoing primary particles in principle so that our adopted value $\dot{N}_{\rm p,out}$ corresponds to the upper limit. On the other hand, the kinetic energy flux of the ingoing primary particles $\dot{N}_{\rm p,in}\gamma_{\rm p}m_{\rm e}c^2$ is constrained by the observed thermal luminosity $L_{\rm th}$, 
\begin{eqnarray}\label{sec2:dotN_p,in}
\dot{N}_{\rm p,in}=\frac{L_{\rm th}}{\gamma_{\rm p}m_{\rm e}c^2}.
\end{eqnarray}
These are the number fluxes of primary particles flowing the open zone of the magnetosphere. In other word, these number fluxes correspond to the total number of particles passing through the area $\pi r_{\rm pc}^2(r/R_{\rm NS})^3$ in unit time at the location $r$, where $r_{\rm pc}=R_{\rm NS}^{3/2}R_{\rm lc}^{-1/2}$ is the polar cap radius. To estimate the number of primary particles $N_{\rm p}(r)$ from $\dot{N}_{\rm p,out}$ or $\dot{N}_{\rm p,in}$, we need to specify the characteristic timescale at $r$. Since our aim is the estimation of the non-thermal X-ray luminosity, we always consider the advection timescale of non-thermal X-ray-emitting particles, $t_{\rm ad} (= r_{\rm sec}/c)$ as the characteristic timescale at $r_{\rm sec}$. Therefore, we estimate the number $N_{\rm p}(r_{\rm sec})$ using the relation $N_{\rm p}(r_{\rm sec})\sim \dot{N}_{\rm p,out}t_{\rm ad}(r_{\rm sec})$ or $N_{\rm p}(r_{\rm sec})\sim \dot{N}_{\rm p,in}t_{\rm ad}(r_{\rm sec})$. In this case, the volume in which the number $N_{\rm p}(r)$ is counted corresponds to $\sim r\times\pi r_{\rm pc}^2(r/R_{\rm NS})^3$. Since we focus on the particles which emit at the observed frequency $\nu_{\rm obs}$, we use $\gamma_{\rm s, syn}(r_{\rm sec})$ or $\gamma_{\rm s, spa}(r_{\rm sec})$ as the Lorentz factor of the secondary particles to calculate the number of the primary particles $N_{\rm p}(r_{\rm sec})$. However, the number $\dot{N}_{\rm p}t_{\rm ad}(t_{\rm sec})$ is implicitly assumed that the synchrotron cooling timescale $t_{\rm cool}(r_{\rm sec})$ for the secondary particles with the Lorentz factor $\gamma_{\rm s,syn}(r_{\rm sec})$ or $\gamma_{\rm s, spa}(r_{\rm sec})$ is larger than the timescale $t_{\rm ad}(r_{\rm sec})$. The cooling timescale is given by
\begin{eqnarray}\label{sec2:t_cool}
t_{\rm cool}(r_{\rm sec})\sim\left\{ \begin{array}{ll}
{\displaystyle \frac{\gamma_{\rm s,syn}(r_{\rm sec})\alpha(r_{\rm sec})m_{\rm e}c^2}{P_{\rm syn}(r_{\rm sec})}} & [\nu_{\rm obs} > \nu_{\rm ct}(r_{\rm sec})] \\
 & \\
{\displaystyle \frac{\gamma_{\rm s,spa}(r_{\rm sec})\alpha(r_{\rm sec})m_{\rm e}c^2}{P_{\rm syn}(r_{\rm sec})}} & [\nu_{\rm obs} < \nu_{\rm ct}(r_{\rm sec})]. \\
\end{array} \right.
\end{eqnarray}
In the case of the condition $t_{\rm cool}(r_{\rm sec})<t_{\rm ad}(r_{\rm sec})$, we should consider the effect of the synchrotron cooling. Taking account for both cases, we evaluate the effective number of primary particles as
\begin{eqnarray}\label{sec2:N_p}
N_{\rm p}(r_{\rm sec})\sim\min\{t_{\rm cool}(r_{\rm sec}), t_{\rm ad}(r_{\rm sec})\}\times\left\{ \begin{array}{ll}
\dot{N}_{\rm p,out} & ({\rm outward}), \\ 
\dot{N}_{\rm p,in} & ({\rm inward}). \\
\end{array} \right. 
\end{eqnarray}
The effective number $N_{\rm p}(r_{\rm sec})$ is inversely proportional to the Lorentz factor $\gamma_{\rm p}$.

Using equations (\ref{sec2:tau_Bgamma}), (\ref{sec2:tau_gammagamma}), (\ref{sec2:N_gamma}) and (\ref{sec2:N_p}), we obtain the non-thermal luminosity from the secondary particles in our model. For the photon-photon pair production case, the required condition for the non-thermal X-ray luminosity becomes
\begin{eqnarray}\label{sec2:number}
P_{\rm syn}(r_{\rm sec})N_{\gamma}(r_{\rm pri},r_{\rm sec})\tau(r_{\rm sec})N_{\rm p}(r_{\rm sec})>L_{\rm syn}.
\end{eqnarray}
This condition gives both the lower and the upper limits for the emission location. Here, the dependences of the number $N_{\gamma}(r_{\rm pri},r_{\rm sec})$ and the number flux $\dot{N}_{\rm p}$ on the Lorentz factor $\gamma_{\rm p}$ are given as $N_{\gamma}(r_{\rm pri},r_{\rm sec})\propto\gamma_{\rm p}$ and $\dot{N}_{\rm p}\propto\gamma_{\rm p}^{-1}$. Then, inequality (\ref{sec2:number}) does not depend on the assumed Lorentz factor of the primary particles $\gamma_{\rm p}$ in equation (\ref{sec2:gamma_p}). It should be noted that all constraints which we introduce in section 2.2 always give the most optimistic limits for the emission regions (i.e., the value of lower limit is the smallest one and that of the upper limit is the largest one) when we use the largest value of $\gamma_{\rm p}$ as equation (\ref{sec2:gamma_p}) in our model.

For the case of the magnetic pair production, the pair cascade is very effective. In the view of the particle energy conservation, the larger number of the particles with the small Lorentz factor [compared with the initial Lorentz factor $\gamma_{\rm s,pair}(r_{\rm pri})$] would be produced than the number of them with the Lorentz factor $\gamma_{\rm s,pair}(r_{\rm pri})$ during the pair cascade. The lower threshold value of the Lorentz factor for the magnetic pair production $\gamma_{\rm s,lt}(r_{\rm sec})$ at the distance $r_{\rm sec}$ is described by using inequality (\ref{sec2:Bcreation}) as
\begin{eqnarray}\label{sec2:gamma_s,lt}
\gamma_{\rm s,lt}(r_{\rm sec})=\frac{1}{15}\frac{B_{\rm q}}{B_{\perp}(r_{\rm sec})}.
\end{eqnarray} 
From the conservation of the particle energy, the number of produced particles is maximally increased by a factor of $\gamma_{\rm s,pair}(r_{\rm pri})/\gamma_{\rm s,lt}(r_{\rm sec})$ so that we modify inequality (\ref{sec2:number}) as
\begin{eqnarray}\label{sec2:number2}
P_{\rm syn}(r_{\rm sec})N_{\gamma}(r_{\rm pri},r_{\rm sec})\tau(r_{\rm sec})N_{\rm p}(r_{\rm sec})\frac{\gamma_{\rm s,pair}(r_{\rm pri})}{\gamma_{\rm s,lt}(r_{\rm sec})}>L_{\rm syn},
\end{eqnarray}
for the case of the magnetic pair production. Because of the difference of the dependence on the distance $r_{\rm sec}$ for the Lorentz factors $\gamma_{\rm s,syn}(r_{\rm sec})$ and $\gamma_{\rm s,spa}(r_{\rm sec})$, inequality (\ref{sec2:number2}) gives the upper limit of the emission region for the conventional synchrotron radiation and the lower limit of the emission region for the small-pitch-angle one. Table 2 summarizes the definitions of our used limits for locations of emission region.

\section{OUTGOING PRIMARY PARTICLES}
In this section, we consider the synchrotron radiation originated from the secondary particles produced by the outgoing primary particles. Our model for the case of the outgoing primary particles is the one-zone approximation ($r_{\rm pri}\sim r_{\rm sec}$) which the locations of the particle acceleration, the $\gamma$-ray emission, the pair production and the non-thermal X-ray emission are almost the same one. We omit subscripts ``pri'' and ``sec'' in this section. For the dipole magnetic field, we evaluate the pitch angle $\alpha(r)\sim\alpha_{\ast}(r/R_{\rm lc})^{1/2}$ with $\alpha_{\ast}<1$ as a first approximation (e.g., Tang et al. 2008). In this approximation, we consider close-up of two field lines, which approximate two concentric circles with curvature radii $R_{\rm cur}(r)$ and $R_{\rm cur}(r)+h_{\rm m}(r)$, where $h_{\rm m}(r)$ denotes the length between two field lines. We introduce a dimensionless parameter $f(r)=h_{\rm m}(r)/R_{\rm lc}$ which is approximated by $f(r)=f(R_{\rm lc})(r/R_{\rm lc})^{3/2}$ considering the conservation of magnetic flux. The pitch angle $\alpha$ comes from the collision angle of curvature photons emitted at inner field line with the outer field line. We approximately obtain the relation $\alpha^2(r)\sim 2h_{\rm m}(r)/R_{\rm cur}(r)\sim 2f(R_{\rm lc})(r/R_{\rm lc})$ using $R_{\rm cur}\sim(rR_{\rm lc})^{1/2}$. A constant parameter $\alpha_{\ast}$ corresponds to the pitch angle at the light cylinder, $\alpha_{\ast}\equiv\alpha(R_{\rm lc})$. In the following, we evaluate inequalities (\ref{sec2:energy}) and (\ref{sec2:frequency}), which do not depend on the pair production processes. In sections 3.1 and 3.2, we describe inequalities (\ref{sec2:Bcreation}), (\ref{sec2:gcreation}), (\ref{sec2:number}) and (\ref{sec2:number2}) depending on the pair production mechanisms.

First, we evaluate the constraint for the energy of the secondary particles, inequality (\ref{sec2:energy}). The Lorentz factor $\gamma_{\rm s,pair}$ is evaluated by using equations (\ref{sec2:E_cur}), (\ref{sec2:gamma_p}) and (\ref{sec2:gamma_s}) as
\begin{eqnarray}\label{sec3:gamma_s}
\gamma_{\rm s,pair}\sim2.6\times10^2P_0^{-13/2}B_{\rm s,12}^3r_6^{-1/2}.
\end{eqnarray}
Substituting equations (\ref{sec2:gamma_s}), (\ref{sec2:nu_syn}) and (\ref{sec2:nu_spa}) into inequality (\ref{sec2:energy}), we obtain the upper limit for the emission locations $r_{\gamma{\rm syn}}$ and $r_{\gamma{\rm spa}}$ as 
\begin{eqnarray}\label{sec3:r_gamma}
\begin{array}{ll}
r_{\gamma{\rm syn},6}\sim 11\alpha_{\ast}^{2/7}\nu_{\rm obs,keV}^{-2/7}P_0^{-27/7}B_{\rm s,12}^2 & (\nu_{\rm obs}>\nu_{\rm ct}) \\
 & \\
r_{\gamma{\rm spa},6}\sim 12\nu_{\rm obs,keV}^{-2/7}P_0^{-13/7}B_{\rm s,12}^{8/7} & (\nu_{\rm obs}<\nu_{\rm ct}). \\
\end{array} 
\end{eqnarray}

From the constraint of the characteristic frequency in the conventional synchrotron radiation, inequality (\ref{sec2:frequency}), we obtain the lower limits for the emission region $r_{\rm ct}$ as 
\begin{eqnarray}\label{sec3:r_ct}
r_{\rm ct,6}\sim6.8\alpha_{\ast}^{-2/7}\nu_{\rm obs,17}^{-2/7}P_0^{1/7}B_{\rm s,12}^{2/7}.
\end{eqnarray}
For the region $r > r_{\rm ct}~(\nu_{\rm obs} > \nu_{\rm ct})$, we consider the conventional synchrotron radiation. However, for the region $r<r_{\rm ct}~(\nu_{\rm obs}<\nu_{\rm ct})$, we should consider the small-pitch-angle synchrotron radiation. In this case, the constraint of the characteristic frequency, inequality (\ref{sec2:frequency2}), is given by
\begin{eqnarray}\label{sec3:r_cut}
r_{\rm cut,6}\sim2.9\nu_{\rm obs,keV}^{-1/3}B_{\rm s,12}^{1/3}.
\end{eqnarray}
We discuss two emission mechanisms separately.

\subsection{Death Line for Magnetic Pair Production}

In this subsection we derive the constraints for the particle production threshold of the magnetic pair production, inequality (\ref{sec2:Bcreation}), and for the non-thermal X-ray luminosity, inequality (\ref{sec2:number}), to the emission location. Then, we compare results with the observational values. 

For the constraints of pair production threshold, we substitute equations (\ref{sec2:E_cur}) and (\ref{sec2:mfp}) into inequality (\ref{sec2:Bcreation}) and obtain the upper limit for the emission location $r_{\rm B\gamma}$ as
\begin{eqnarray}\label{sec3:r_Bgamma}
r_{\rm B\gamma,6}\sim1.1\alpha_{\ast}^{1/3}P_0^{-7/3}B_{\rm s,12}^{4/3}.
\end{eqnarray}

Since the pair conversion is very efficient for the magnetic pair production, we approximate $\tau_{\rm B\gamma}\sim1$ as long as the pair production condition, inequality (\ref{sec2:Bcreation}), is satisfied. We only consider the case $t_{\rm cool}<t_{\rm ad}$ in inequality (\ref{sec2:number2}) for the magnetic pair production because of the strong magnetic field at $r<r_{\rm B\gamma}$. Then, the conditions of the non-thermal luminosity, inequality (\ref{sec2:number2}), give the upper limit for the emission location $r_{\rm LBsyn}$ at $r > r_{\rm ct}$ and the lower limit for the emission location $r_{\rm LBspa}$ at $r < r_{\rm ct}$. The limits $r_{\rm LBsyn}$ and $r_{\rm LBspa}$ are derived by substituting equations (\ref{sec2:gamma_s}), (\ref{sec2:tau_Bgamma}), (\ref{sec2:N_gamma}), (\ref{sec2:dotN_p,out}), (\ref{sec2:N_p}) and (\ref{sec2:gamma_s,lt}) into inequality (\ref{sec2:number2}) as 
\begin{eqnarray}\label{sec3:r_nB}
r_{\rm LBsyn,6}\sim2.5\times10^{-3}\alpha_{\ast}^2\varepsilon_{\rm syn,-3}^{-4/3}\nu_{\rm obs,keV}^{2/3}P_0^{-31/3}B_{\rm s,12}^{14/3},
\end{eqnarray}
and 
\begin{eqnarray}\label{sec3:r_LBspa}
r_{\rm LBspa,6}\sim7.7\times10^3\alpha_{\ast}^{-2}\varepsilon_{\rm syn,-3}\nu_{\rm obs,keV}^{-1}P_0^8B_{\rm s,12}^{-3},
\end{eqnarray}
where we introduce the ratio of the non-thermal luminosity to the spin-down luminosity, $\varepsilon_{\rm syn}\equiv L_{\rm syn}/L_{\rm sd}$. 

\begin{figure}
 \begin{center}
  \includegraphics[width=100mm, angle=270]{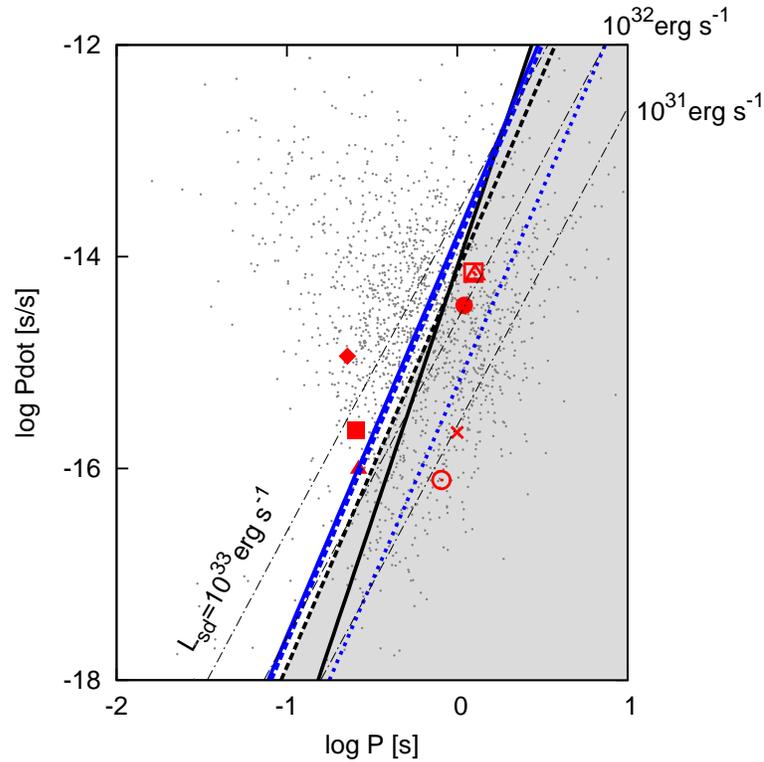}
 \end{center}
 \caption{Synchrotron radiation death lines on the $P$-$\dot{P}$ diagram. Secondary particles are produced by the outgoing curvature photons via the magnetic pair production. The death lines of the conventional synchrotron radiation, which obtained from inequalities (\ref{sec3:outBgammaPdotP}) are denoted as thick black solid ($r_{\rm LBsyn} < r_{\rm B\gamma}$) and dashed lines($r_{\rm LBsyn} > r_{\rm B\gamma}$). The death lines of the small-pitch-angle synchrotron radiation, which obtained from inequalities (\ref{sec3:outBgammaPdotP2}) are denoted as thick blue dotted ($r_{\rm LBspa} < r_{\rm cut}$), solid ($r_{\rm LBspa} > r_{\rm cut}$ and $r_{\rm ct} < r_{\rm B\gamma}$) and dashed lines ($r_{\rm LBspa} > r_{\rm cut}$ and $r_{\rm ct} > r_{\rm B\gamma}$). The thin dot-dashed lines denote the spin-down luminosity $L_{\rm sd}=10^{33}, 10^{32}$ and $10^{31}$ erg s$^{-1}$. Each large point denotes X-ray detected old pulsars, J0108-1431 ($\circ$), B0628-28 ($\Box$), B0834+06 ($\triangle$), B0943+10 ($\bullet$), B0950+08 ($\blacksquare$), B1451-68 ($\blacktriangle$) and B1929+10 ($\blacklozenge$). A large cross denotes the fiducial pulsar. Small dots denote other pulsars taken from ATNF Pulsar Catalog \citep{Ma05}. Four of our samples and a fidutial pulsar locate in the filled region (below the death line for the equalities \ref{sec3:outBgammaPdotP}) so that the model in section 3.1 does not explain their non-thermal X-ray emissions.}
 \label{fig1}
\end{figure}

From equations (\ref{sec3:r_gamma}), (\ref{sec3:r_ct}), (\ref{sec3:r_cut}), (\ref{sec3:r_Bgamma}), (\ref{sec3:r_nB}) and (\ref{sec3:r_LBspa}), we obtain the constraint for the emission location in the case that secondary particles are produced by outgoing curvature photons through the magnetic pair production. In the case of the conventional synchrotron radiation, $\nu_{\rm obs}>\nu_{\rm ct}~(r > r_{\rm ct})$, the obtained constraint is $r_{\rm ct} < r < \min\{r_{\gamma{\rm syn}},r_{\rm B\gamma},r_{\rm LBsyn}\}$ for the emission region. In the case of the small-pitch-angle synchrotron radiation $\nu_{\rm obs}<\nu_{\rm ct}~(r< r_{\rm ct})$, the constraint is $\max\{r_{\rm LBspa}, r_{\rm cut}\}<r<\min\{r_{\gamma{\rm spa}},r_{\rm B\gamma},r_{\rm ct}\}$. Since our samples satisfy the conditions $r_{\rm B\gamma}<r_{\rm \gamma syn}$ and $r_{\rm B\gamma}<r_{\rm \gamma spa}$, we take $\min\{r_{\rm B\gamma}, r_{\gamma{\rm syn}}\}=r_{\rm B\gamma}$ and $\min\{r_{\rm B\gamma}, r_{\gamma{\rm spa}}\}=r_{\rm B\gamma}$. The condition $r_{\rm cut} < r_{\rm ct}$ is always satisfied because of the small value of the pitch angle $\alpha$ at the inner magnetosphere. On the other hand, only two pulsars with small period $(P_0\lesssim0.3)$ and large spin-down luminosity $(L_{\rm sd,31}\lesssim10^{1.5})$ in our samples, B0950+08 and B1929+10, satisfy the conditions $r_{\rm LBsyn}<r_{\rm B\gamma}$ and $r_{\rm LBspa}<r_{\rm cut}$ using observational values in table 1. Therefore, we obtain the constraints for the emission region of old pulsars as
\begin{eqnarray}\label{sec3:outBgamma}
\begin{array}{ll}
r_{\rm ct}<r<\min\{r_{\rm B\gamma},r_{\rm LBsyn}\} & (\nu_{\rm obs}>\nu_{\rm ct}) \\
 & \\
\max\{r_{\rm cut}, r_{\rm LBspa}\}<r<\min\{r_{\rm B\gamma},r_{\rm ct}\}. & (\nu_{\rm obs}<\nu_{\rm ct}). \\
\end{array} 
\end{eqnarray} 
For the conventional synchrotron radiation, the inequality $\min\{r_{\rm B\gamma},r_{\rm LBsyn}\}/r_{\rm ct}>1$ is the condition which secondary pairs produced by the outgoing primary particles through the magnetic pair production reproduce the observed non-thermal X-ray emission. Using this condition, we obtain the death lines for the conventional synchrotron radiation on the $P$ versus $\dot{P}$ plane as
\begin{eqnarray}\label{sec3:outBgammaPdotP}
\dot{P}>\left\{ \begin{array}{ll}
8.1\times10^{-15}\alpha_{\ast}^{-24/23}\varepsilon_{\rm syn,-3}^{14/23}\nu_{\rm obs,keV}^{-10/23}P_0^{110/23}~{\rm s~s}^{-1} & (r_{\rm LBsyn} < r_{\rm B\gamma}). \\
 & \\
7.1\times10^{-15}\alpha_{\ast}^{-13/11}\nu_{\rm obs,keV}^{-6/11}P_0^{41/11}~{\rm s~s}^{-1} & (r_{\rm LBsyn} > r_{\rm B\gamma}) \\
\end{array} \right.
\end{eqnarray}
On the other hand, the inequality $\min\{r_{\rm B\gamma},r_{\rm ct}\}/\max\{r_{\rm LBspa}, r_{\rm cut}\}>1$ is the condition for the small-pitch-angle synchrotron radiation. The death lines are given as
\begin{eqnarray}\label{sec3:outBgammaPdotP2}
\dot{P}>\left\{ \begin{array}{ll}
6.1\times10^{-16}\alpha_{\ast}^{-12/37}\varepsilon_{\rm syn,-3}^{24/37}\nu_{\rm obs,keV}^{-22/37}P_0^{137/37}~{\rm s~s}^{-1} & (r_{\rm LBspa} < r_{\rm cut}) \\
 & \\
1.6\times10^{-14}\alpha_{\ast}^{-24/23}\varepsilon_{\rm syn,-3}^{14/23}\nu_{\rm obs,keV}^{-10/23}P_0^{87/23}~{\rm s~s}^{-1} & (r_{\rm LBspa} > r_{\rm cut}~{\rm and}~r_{\rm ct} < r_{\rm B\gamma}) \\
 & \\
1.3\times10^{-14}\alpha_{\ast}^{-14/13}\varepsilon_{\rm syn,-3}^{6/13}\nu_{\rm obs,keV}^{-6/13}P_0^{49/13}~{\rm s~s}^{-1} & (r_{\rm LBspa} > r_{\rm cut}~{\rm and}~r_{\rm ct}>r_{\rm B\gamma}). \\
\end{array} \right.
\end{eqnarray}
These five death lines with $\alpha_{\ast}=\varepsilon_{\rm syn,-3}=\nu_{\rm obs,keV}=1$ are depicted on the $P$ versus $\dot{P}$ diagram in figure \ref{fig1}. Black lines denote the death lines for the conventional synchrotron radiation (a solid line for $r_{\rm LBsyn} < r_{\rm B\gamma}$ and a dashed line for $r_{\rm LBsyn} > r_{\rm B\gamma}$), and blue lines denote the death lines for the small-pitch-angle synchrotron radiation (a dotted line for $r_{\rm LBspa} < r_{\rm cut}$, a solid line for $r_{\rm LBspa} > r_{\rm cut}$ and $r_{\rm ct} < r_{\rm B\gamma}$, and a dashed line for $r_{\rm LBspa} > r_{\rm cut}$ and $r_{\rm ct} > r_{\rm B\gamma}$). Note that for the small-pitch-angle synchrotron radiation, the allowed region is small compared with the case of the conventional synchrotron radiation due to the large value of the lower limit $r_{\rm LBspa}$. In the case of $t_{\rm cool} < t_{\rm ad}$, the required number of secondary particles with Lorentz factor $\gamma$ are described as $L_{\rm syn}t_{\rm ad}/(\gamma\alpha m_{\rm e} c^2)$. Since $\gamma_{\rm s,spa}\alpha$ is always smaller than $\gamma_{\rm s,syn}\alpha$ for a given frequency, $\nu_{\rm obs}$, the larger number of secondary particles are required in the small-pitch-angle synchrotron radiation. Although the advection timescale is $t_{\rm ad}\propto r$, the value $\gamma_{\rm s,spa}\alpha$ is proportional to $r^{7/2}$ so that the value of the lower limit $r_{\rm LBspa}$ should be large. 

We plot our samples as red points. We also plot a red cross as a fiducial pulsar. We see that pulsars with $L_{\rm sd,31}\lesssim (10-100)$ does not satisfy our constraints (the filled region for the case of the conventional synchrotron radiation). For these pulsars, both the conventional and the small-pitch-angle synchrotron radiations from secondary pairs which are produced from the outgoing $\gamma$-ray photons through the magnetic pair production is not significantly contributed to their observed non-thermal X-ray emission. Pulsars with small period $P_0\lesssim0.3$ and large spin-down luminosity $L_{\rm sd,31}\gtrsim 10$ satisfy inequalities (\ref{sec3:outBgammaPdotP}) and (\ref{sec3:outBgammaPdotP2}). Since both the conventional and the small-pitch angle synchrotron radiations explain their non-thermal X-ray emission, the allowed range of the emission location is $r_{\rm cut} < r < \max\{r_{\rm B\gamma}, r_{\rm LBsyn}\}$ ($2\lesssim r_6\lesssim 4$ for B1451-68 and $2\lesssim r_6\lesssim10$ for B0950+08). Their upper limit is much smaller values than that of the outer gap model. This mean that the particle acceleration region of pulsars with $L_{\rm sd,31}\lesssim100$ moves to inner magnetosphere ($r_6\lesssim10$) if this model of magnetic pair production is correct for non-thermal emission. For B1929+10 ($L_{\rm sd,31}\sim440$), the upper limit of the emission location are $r_6\lesssim 40$ which is also a bit small for the outer gap model.

\subsection{Death Line for Photon-Photon Pair production}

In the case of the magnetic pair production, since the strong magnetic field is required to produce pairs, the emission region locates near the neutron star surface. On the other hand, the photon-photon process occurs up to relatively far region from the star. Here, we consider the secondary particles produced through the photon-photon pair production. 

First, we consider the threshold of the particle production for the photon-photon process. Using the assumption of the dipole magnetic field, the collision angle between a $\gamma$-ray photon emitted from the outgoing primary particles and a X-ray photon from the polar cap is approximated by
\begin{eqnarray}\label{sec3:theta_col}
1-\cos\theta_{\rm col}\sim\frac{1}{2}\left(\frac{r}{R_{\rm cur}}\right)^2.
\end{eqnarray}
Because of the curved magnetic field line, the collision angle increases with the distance from the neutron star. Substituting equation (\ref{sec3:theta_col}) into inequality (\ref{sec2:gcreation}), we obtain the lower limit for the location of the photon-photon pair production as
\begin{eqnarray}\label{sec3:r_gammagamma}
r_{\gamma\gamma,6}\sim1.5\times10^9T_{\rm pc,6}^{-2}P_0^{15}B_{\rm s,12}^{-6}.
\end{eqnarray}
Since the collision angle becomes smaller as the emission region comes close to the star, the pair production condition gives the lower limit for the emission region in the case of the outgoing primary particles. For a fiducial pulsar ($P_0=B_{\rm s,12}=1$), the value of the lower limit $r_{\gamma\gamma}$ is much larger than the radius of the light cylinder, i.e., the photon-photon pair production is not allowed within the entire magnetosphere. However, because of the strong dependence of the lower limit $r_{\gamma\gamma}$ on the period $P$, the value of $r_{\gamma\gamma}$ for a pulsar with $P_0=0.1$ is only $r_{\gamma\gamma}\sim1.5$ cm, well within the radius of the light cylinder. 

Next, we consider the condition for the non-thermal X-ray luminosity. The optical depth $\tau_{\gamma\gamma}$ is obtained by substituting equation (\ref{sec3:theta_col}) into equation (\ref{sec2:tau_gammagamma}) as
\begin{eqnarray}\label{sec3:tau_gammagamma}
\tau_{\gamma\gamma}\sim9.5\times10^{-10}L_{\rm th,28}T_{\rm pc,6}^{-1}P_0^{-1}.
\end{eqnarray}
We consider the case of the conventional synchrotron radiation ($r > r_{\rm ct}$). Substituting equations (\ref{sec2:N_s}), (\ref{sec2:N_gamma}), (\ref{sec2:dotN_p,out}), (\ref{sec2:N_p}), (\ref{sec2:t_cool}) and (\ref{sec3:tau_gammagamma}) into inequality (\ref{sec2:number}), we obtain the lower limit $r_{\rm L\gamma syn1}~(t_{\rm ad}>t_{\rm cool})$ and the upper limit $r_{\rm L\gamma syn2}~(t_{\rm ad}<t_{\rm cool})$ for the emission region as
\begin{eqnarray}\label{sec3:r_ngamma1}
r_{\rm L\gamma syn1,6}\sim8.4\times10^4\alpha_{\ast}^{-2/9}\xi^{4/9}L_{\rm sd,31}^{-4/9}\nu_{\rm obs,keV}^{-2/9}T_{\rm pc,6}^{4/9}P_0^{7/9}B_{\rm s,12}^{2/9},~~(t_{\rm ad} > t_{\rm cool})
\end{eqnarray}
and
\begin{eqnarray}\label{sec3:r_ngamma2}
r_{\rm L\gamma syn2,6}\sim2.9\times10^{-2}\alpha_{\ast}\xi^{-1}L_{\rm sd,31}\nu_{\rm obs,keV}T_{\rm pc,6}^{-1}P_0^{-2}B_{\rm s,12},~~(t_{\rm ad} < t_{\rm cool})
\end{eqnarray}
where $\xi=L_{\rm syn}/L_{\rm th}$. In the region close to the star, X-ray emitting particles are required to have the small value of Lorentz factor because of the large magnitude of the magnetic field. Then, the energy of a single emitting particle is much small and the large number of pairs is required to explain the observed luminosity of the non-thermal X-ray component. Therefore, we obtain the lower limit $r_{\rm n\gamma1}$ for the emission location from inequality (\ref{sec2:number}) in the case $t_{\rm cool}< t_{\rm ad}$. On the other hand, in the far region from the star, the magnitude of the magnetic field is weak. Then, the emissivity of the synchrotron radiation decreases so that the large number of pairs is also required. In this reason, we obtain the upper limit $r_{\rm n\gamma2}$ for the emission region by the condition of the non-thermal X-ray luminosity in the case, $t_{\rm cool}>t_{\rm ad}$. 

Note that in our one-zone approximation ($r_{\rm pri}\sim r_{\rm sec}$), it seems to underestimate the right-hand side of inequality (\ref{sec2:number}) for the case, $t_{\rm ad}<t_{\rm cool}$, because secondary particles do not significantly lose their energy at the region $r_{\rm sec}$. However, since the magnitude of the magnetic field strongly depends on the distance from the centre of the neutron star, the characteristic frequency and the emissivity of the synchrotron radiation steeply decrease at the outer region from the emission region, $r$. Therefore, we consider that our estimated luminosity does not give significantly underestimated value in our model. 

The obtained constraints for the emission region are summarized as the condition $\max\{r_{\rm ct},r_{\rm L\gamma syn1},r_{\gamma\gamma}\}<r<\min\{r_{\gamma{\rm syn}},r_{\rm L\gamma syn2}\}$. For the old pulsars, the relations $r_{\rm L\gamma syn1}>r_{\rm ct}$ and $r_{\gamma{\rm syn}}>r_{\rm L\gamma syn2}$ are satisfied. Therefore, we reduce the obtained condition for the emission region as 
\begin{eqnarray}
\max\{r_{\gamma\gamma},r_{\rm L\gamma syn1}\}<r<r_{\rm L\gamma syn2}.
\end{eqnarray}

\begin{figure}
 \begin{center}
  \includegraphics[width=100mm, angle=270]{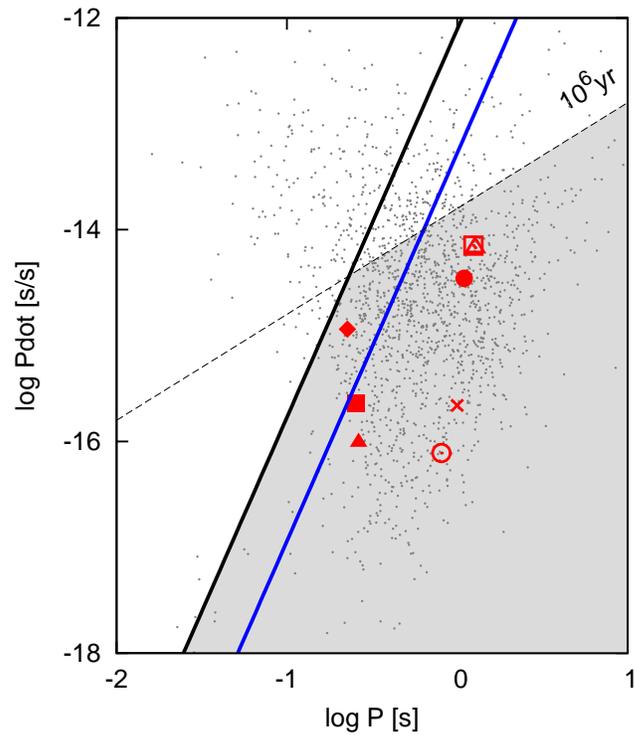}
 \end{center}
 \caption{Synchrotron radiation death lines on the $P$-$\dot{P}$ diagram. Secondary particles are produced by the outgoing curvature photons via photon-photon pair production. The black and blue solid lines denote the death lines obtained from inequalities (\ref{sec3:outgammagammaPdotP}) for $r_{\gamma\gamma} < r_{\rm L\gamma syn1}$ and $r_{\gamma\gamma}>r_{\rm L\gamma syn1}$, respectively. Thin dashed line denotes the characteristic age $10^6$yr. Each point is same as in figure \ref{fig1}. Our samples locate the filled region (below the death line for inequality (\ref{sec3:outgammagammaPdotP}) with the condition $r_{\gamma\gamma} < r_{\rm L\gamma syn1}$) so that the model in section 3.2 does not explain their non-thermal X-ray emissions.} 
 \label{fig3}
\end{figure}

In figure \ref{fig3}, we plot the conditions $r_{\rm L\gamma syn2}/\max\{r_{\gamma\gamma},r_{\rm L\gamma syn1}\}>1 $ given by 
\begin{eqnarray}\label{sec3:outgammagammaPdotP}
\dot{P}>\left\{ \begin{array}{ll}
7.8\times10^{-13}\alpha_{\ast}^{-22/33}\xi^{26/33}\nu_{\rm obs,keV}^{-22/33}T_{\rm pc,6}^{26/33}P^{121/33}~{\rm s~s}^{-1} & (r_{\gamma\gamma} < r_{\rm L\gamma syn1}), \\
 & \\
5.3\times10^{-14}\alpha_{\ast}^{-2/9}\xi^{2/9}\nu_{\rm obs,keV}^{-2/9}T_{\rm pc,6}^{-2/9}P^{11/3}~{\rm s~s}^{-1}  & (r_{\gamma\gamma} > r_{\rm L\gamma syn1}). \\
\end{array} \right.
\end{eqnarray}
with $\alpha_{\ast}=\xi=\nu_{\rm obs,keV}=T_{\rm pc,6}=1$ as black ($r_{\gamma\gamma} < r_{\rm L\gamma syn1}$) and blue ($r_{\gamma\gamma} > r_{\rm L\gamma syn1}$) solid lines, respectively. Here, we use the relation $L_{\rm sd}=4\pi^2 I\dot{P}P^{-3}$. The filled region in figure \ref{fig3} shows the parameter space of pulsars whose synchrotron radiation does not explain the observed non-thermal X-ray emission because of the relation $r_{\rm L\gamma syn1}>r_{\rm L\gamma syn2}$. Since the inequality $r_{\gamma\gamma}<r_{\rm L\gamma syn1}$ is satisfied for B1929+10, we see that none of our samples satisfy inequality (\ref{sec3:outgammagammaPdotP}). Note that we do not account for the bulk neutron star surface emission as seed photons so that our obtained constraints are only valid for pulsars with $\gtrsim 1$Myr. We depict a thin dashed line which shows the characteristic age 1Myr. 

\begin{figure}
 \begin{center}
  \includegraphics[width=100mm, angle=270]{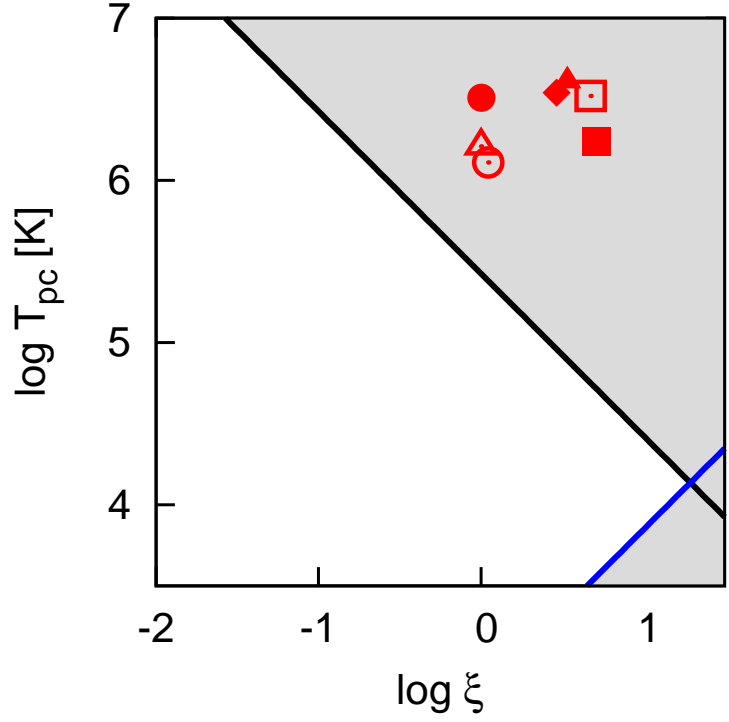}
 \end{center}
 \caption{The luminosity ratio $\xi$ vs. the temperature of the polar cap $T_{\rm pc}$ plane. The solid lines denote the death lines for B1929+10 obtained from inequalities (\ref{sec3:outgammagammaPdotP}) for $r_{\gamma\gamma} < r_{\rm L\gamma syn1}$  (black) and $r_{\gamma\gamma}>r_{\rm L\gamma syn1}$ (blue). Each point is same as in figure \ref{fig1}. The filled region denotes where a pulsar with same $P$ and $\dot{P}$ of B1929+10 does not satisfy the condition $r_{\rm n\gamma2}/\max\{r_{\gamma\gamma}, r_{\rm n\gamma1}\}>1$. Pulsars with the ratio $\xi>1$ and the temperature $T_{\rm pc,6}>1$ are excluded even if period and its time derivative are as same as that of B1929+10.} 
 \label{fig4}
\end{figure}

In order to clarify the dependence of inequality (\ref{sec3:outgammagammaPdotP}) on $\xi$ and $T_{\rm pc}$, we also show the $\xi$-$T_{\rm pc}$ diagram in figure \ref{fig4}. As a representative of our samples, the solid lines denote the death lines for $\xi$ and $T_{\rm pc}$ in the case of B1929+10. The meaning of the filled region in figure \ref{fig4} is the same as in figure \ref{fig3}. The death lines in the case of other samples give more strict upper limits for $\xi$. In our samples, the temperature of the heated polar cap is $T_{\rm pc}\gtrsim10^6$K. Then, the luminosity of the synchrotron radiation have to be smaller than the thermal one ($\xi\lesssim1$). Therefore, in the case of the photon-photon pair production, the synchrotron radiation does not explain the luminosity of non-thermal component, which is larger than the thermal luminosity from the heated polar cap region.

Note that if we consider the small-pitch angle synchrotron radiation, the lower limit $r_{\rm L\gamma spa1}$ ($t_{\rm ad} > t_{\rm cool}$) from inequality (\ref{sec2:number}) is obtained as 
\begin{eqnarray}\label{sec3:r_ngammaspa1}
r_{\rm L\gamma spa1,6}\sim1.8\times10^3\alpha_{\ast}^{-1/4}\xi^{1/4}L_{\rm sd,31}\nu_{\rm obs,keV}^{-1/4}T_{\rm pc,6}^{1/4}P_0^{1/2}B_{\rm s,12}^{1/4},~~(t_{\rm ad} < t_{\rm cool})
\end{eqnarray}
Then, one of the condition $r_{\rm L\gamma spa1} < r_{\rm ct}$ is given as
\begin{eqnarray}\label{sec3:outgammagammaPdotP2}
\dot{P}>
2.6\times10^{-7}\alpha_{\ast}^{2/15}\xi^{14/15}\nu_{\rm obs,keV}^{2/15}T_{\rm pc,6}^{14/15}P^{61/15}~{\rm s~s}^{-1}.
\end{eqnarray}
Clearly, old pulsars do not satisfy this condition. Actually, we only consider second-generation particles as non-thermal X-ray-emitting particles because of small optical depth for the photon-photon production process. In this case, perpendicular momentum of these particles is initially relativistic and subsequent small-pitch-angle emission (after losing perpendicular momentum) will be negligible compared to the conventional synchrotron radiation. We conclude that the synchrotron radiation from secondary pairs produced by the photon-photon pair production does not explain the observed non-thermal emission from old pulsars. 

\section{INGOING PRIMARY PARTICLES}

In previous section, we show that the observed non-thermal X-ray emission is not reproduced by our model which secondary particles are produced by the outgoing primary particles. One of main problems for the outgoing case is that the value of the pitch angle remains small ($\sim10^{-2}$) near the neutron star surface. The small value of the pitch angle decreases the efficiency of the particle production. As a result, the upper limits for the emission region, $r_{\rm B\gamma}$, $r_{\rm LB syn}$ and $r_{\rm L\gamma syn2}$ are much small values. On the other hand, if we consider the ingoing primary particles, curvature photons from them cross the magnetic field line almost perpendicular in principle (figure 3 in Wang et al. 1998; see also figure \ref{figA1}). Then, the value of the pitch angle no longer represents the function of distance $r$ as $\alpha\sim\alpha_{\ast}(r/R_{\rm lc})^{1/2}$ used in the outgoing case. We set the pitch angle $\alpha$ as a free parameter of an order unity for the ingoing case. We only consider the conventional synchrotron radiation since the two frequencies is nearly same values, $\nu_{\rm ct}\sim\nu_{\rm cut}$ $(r_{\rm ct}\sim r_{\rm cut})$ for the pitch angle $\alpha\sim1$.

Here, we should consider the curvature radiation region from primary particles $r_{\rm pri}$ and the synchrotron radiation region from secondary particles $r_{\rm sec}$ separately. We assume that the point $r_{\rm sec}$ locates at the minimum distance from the centre of the neutron star to the trajectory of the curvature photons emitted from the distance $r_{\rm pri}$ (figure \ref{figA1}). The relation between two points is given in appendix A as
\begin{eqnarray}\label{sec4:r_pri}
r_{\rm pri,6}\sim27r_{\rm sec,6}^{2/3}R_{\rm open,lc}^{1/3}P_0^{1/3},
\end{eqnarray}
where $R_{\rm open,lc}\equiv R_{\rm open}/R_{\rm lc}$ and $R_{\rm open}$ is the maximum distance from the centre of the neutron star to the top of the magnetic loop on a given field line (equation \ref{app:R_open}). We use equation (\ref{sec4:r_pri}) to convert from the location $r_{\rm pri}$ to the location $r_{\rm sec}$ and only consider the constraints on the location $r_{\rm sec}$ in this section. Note that the inequality $r_{\rm pri,6}\lesssim27R_{\rm open,lc}^{1/3}P_0^{1/3}$ corresponds to the inequality $r_{\rm sec}<R_{\rm NS}$. Then, curvature photons emitted from the location $r_{\rm pri}$ collide with the neutron star. In this reason, equation (\ref{sec4:r_pri}) with the condition $r_{\rm sec,6}=1$ gives the lower limit for the location $r_{\rm pri}$. 

The condition of the energy of the secondary particles is $\gamma_{\rm s,pair}(r_{\rm pri}) > \gamma_{\rm s,syn}(r_{\rm sec})$. Using inequality (\ref{sec2:energy}) and equation (\ref{sec4:r_pri}), we obtain the upper limit for the location $r_{\rm sec}$ given by 
\begin{eqnarray}\label{sec4:r_gamma}
r_{\gamma{\rm syn},6}\sim18\alpha^{3/11}\nu_{\rm obs,keV}^{-3/11}R_{\rm open,lc}^{-1/11}P_0^{-40/11}B_{\rm s,12}^{21/11}.
\end{eqnarray}
Note that the obtained value is larger than that of equation (\ref{sec3:r_gamma}). The constraint from the characteristic frequency gives the lower limit for the location $r_{\rm sec}$ as
\begin{eqnarray}\label{sec4:r_ct}
r_{\rm ct,6}\sim2.9\alpha^{-1/3}\nu_{\rm obs,keV}^{-1/3}B_{\rm s,12}^{1/3}.
\end{eqnarray}
Since we do not consider the small-pitch-angle synchrotron radiation, we only use the lower limit $r_{\rm ct}$ as the constraint from the characteristic frequency.

There is an another difference between the outgoing and the ingoing cases. In the outgoing case, since there is no observational constraint for the number flux of primary particles, we use the spin-down luminosity for the constraint to the particle energy flux. On the other hand, as already mentioned in section 2, the kinetic energy flux of ingoing primary particles is limited by the observed luminosity of the thermal component from the heated polar cap. This modification decreases the number flux of the primary particles (equation \ref{sec2:dotN_p,in}).

\subsection{Death Line for Magnetic Pair Production}

We consider the synchrotron radiation from secondary particles which are produced through the magnetic pair production. In the case of ingoing primary particles, the condition of the magnetic pair production gives the upper limit for the location $r_{\rm sec}$ as 
\begin{eqnarray}\label{sec4:r_Bgamma}
r_{\rm B\gamma,6}\sim2.3\alpha^{3/10}R_{\rm open,lc}^{-1/20}P_0^{-2}B_{\rm s,12}^{6/5}.
\end{eqnarray}

Next, we consider the condition of the non-thermal X-ray luminosity. For the optical depth of the magnetic pair production, we also use $\tau_{\rm B\gamma}\sim1$ as long as the distance to the synchrotron radiation region $r_{\rm sec}$ is smaller than the upper limit, $r_{\rm B\gamma}$. We consider that the value of the pitch angle is an order of unity, $\alpha\sim1$, so that the timescale $t_{\rm cool}$ is always shorter than the timescale $t_{\rm ad}$. We obtain the upper limit for the distance $r_{\rm sec}$ substituting equations (\ref{sec2:gamma_s}), (\ref{sec2:N_s}), (\ref{sec2:N_gamma}), (\ref{sec2:dotN_p,in}) and (\ref{sec2:gamma_s,lt}) into inequality (\ref{sec2:number2}) as
\begin{eqnarray}\label{sec4:r_nB}
r_{\rm LBsyn,6}\sim2.9\times10^{-4}\alpha^{9/7}\xi^{-6/7}\nu_{\rm obs,keV}^{3/7}R_{\rm open,lc}^{-2/7}P_0^{-44/7}B_{\rm s,12}^{3}.
\end{eqnarray}
This value is smaller than that of the outgoing case, equation (\ref{sec3:r_nB}). Since the dependence of the ratio $\xi$ on the upper limit $r_{\rm LBsyn}$ is relatively strong except for the period $P$ and the magnitude of the surface magnetic field $B_{\rm s}$, the constraint from thermal luminosity plays a significant role to limit the emission region.

\begin{figure}
 \begin{center}
  \includegraphics[width=100mm, angle=270]{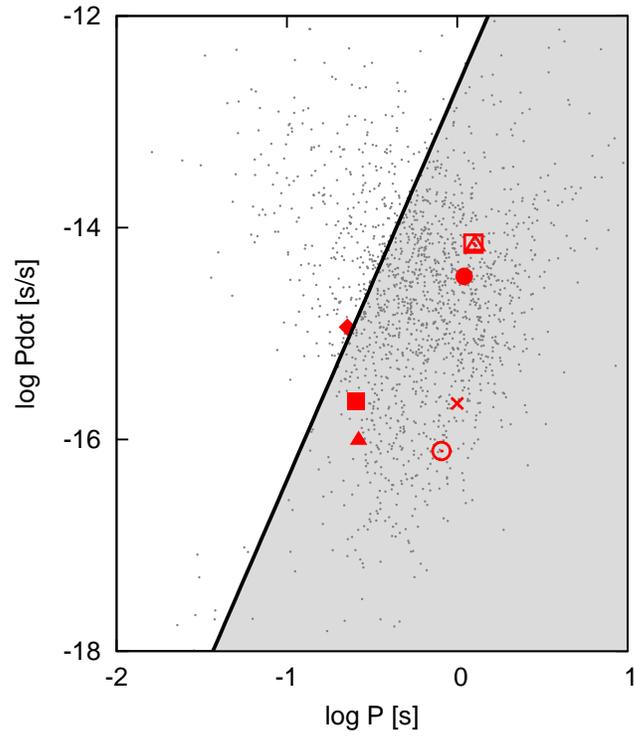}
 \end{center}
 \caption{Synchrotron radiation death line on the $P$-$\dot{P}$ diagram. Secondary particles are produced by the ingoing curvature photons via the magnetic pair production. The solid line denotes the death line obtained from inequality (\ref{sec4:inBgammaPdotP}). Each point is same as in figure \ref{fig1}. Our samples except for B1929+10 locate the filled region (below the death line so that the model in section 4.1 does not explain their non-thermal X-ray emissions.} 
 \label{fig6}
\end{figure}

From equations (\ref{sec4:r_gamma}), (\ref{sec4:r_ct}), (\ref{sec4:r_Bgamma}) and (\ref{sec4:r_nB}), we obtain the constraint $r_{\rm ct}<r_{\rm sec}<\min\{r_{\gamma{\rm syn}},r_{\rm B\gamma},r_{\rm LBsyn}\}$ for the synchrotron radiation region in the case of magnetic pair production. Since our samples satisfy the condition $r_{\rm LBsyn}<r_{\rm B\gamma}<r_{\gamma{\rm syn}}$, we reduce the constraint for the synchrotron radiation region as 
\begin{eqnarray}\label{sec4:inBgamma}
r_{\rm ct}<r_{\rm sec}<r_{\rm LBsyn}.
\end{eqnarray}
Setting $r_{\rm LBsyn}/r_{\rm ct}>1$, we obtain the death line for the ingoing primary particles through the magnetic pair production as
\begin{eqnarray}\label{sec4:inBgammaPdotP}
\dot{P}>2.1\times10^{-13}\alpha^{-17/14}\xi^{9/14}\nu_{\rm obs,keV}^{-4/7}R_{\rm open,lc}^{3/14}P_0^{26/7}~{\rm s~s}^{-1}.
\end{eqnarray}
This relation with $\alpha=\xi=\nu_{\rm obs,keV}=R_{\rm open,lc}=1$ is depicted by a black line in figure \ref{fig6}. The filled region shows the parameter region where the observed non-thermal X-ray emissions are not explained by this model. We obviously see that this scenario does not reproduce the observed non-thermal X-ray emission for our samples except for the pulsar with the largest spin-down luminosity in our samples, B1929+10. The difference between the outward (inequality \ref{sec3:outBgammaPdotP}) and the inward (inequality \ref{sec4:inBgammaPdotP}) cases is the dependence on the thermal luminosity $L_{\rm th}$ (included in the ratio $\xi$). For the ingoing case, since the thermal luminosity reflects the number flux of the primary particles, the small value of the thermal luminosity $L_{\rm th}$ compared with the spin-down luminosity $L_{\rm sd}$ strongly limits the effective number of secondary particles. We conclude that previously suggested synchrotron models \citep{ZC97, WRHZ98} explain the observed non-thermal X-ray component for only a part of old pulsars.

\subsection{Death Line for Photon-Photon Pair Production}

Next, we consider the photon-photon pair production process. We assume that the collision angle between a curvature photon and a thermal photon is $\cos\theta_{\rm col}\sim-1$. It is expected that the optical depth for the photon-photon process $\tau_{\gamma\gamma}$ becomes much larger than that of the model discussed in section 3.2 (equation \ref{sec3:tau_gammagamma}). 

Unlike the outgoing case, the particle production condition (inequality \ref{sec2:gcreation}) gives the upper limit for the emission region due to the following reason. We assume that the collision angle $\theta_{\rm col}$ does not depend on the distance $r_{\rm sec}$. Then, the quantity which depends on the distance $r_{\rm sec}$ is only the energy of the curvature photons in the condition of particle threshold (inequality \ref{sec2:gcreation}) as $E_{\rm cur}\propto r_{\rm pri}^{-1/2}\propto r_{\rm sec}^{-1/3}$. Since the curvature radius of a given field line becomes large as the increase of the distance $r_{\rm sec}$, the curvature photon energy becomes soft for large distance $r_{\rm sec}$. The energy of the seed photons does not depend on the distance $r_{\rm sec}$. Therefore, the condition of the pair production (inequality \ref{sec2:gcreation}) gives the upper limit for the emission region $r_{\rm sec}$. This upper limit is described as,
\begin{eqnarray}\label{sec4:r_gammagamma}
r_{\gamma\gamma,6}\sim8.8\times10^{-4}R_{\rm open,lc}^{-1/2}T_{\rm pc,6}^3P_0^{-20}B_{\rm s,12}^9.
\end{eqnarray}

We consider the condition for the non-thermal X-ray luminosity. Substituting equations (\ref{sec2:N_s}), (\ref{sec2:tau_gammagamma}), (\ref{sec2:N_gamma}) and (\ref{sec2:dotN_p,in}) into inequality (\ref{sec2:number}), we obtain the lower ($r_{\rm L\gamma syn1}$) and the upper ($r_{\rm L\gamma syn2}$) limits for the location $r_{\rm sec}$ as 
\begin{eqnarray}\label{sec4:r_ngamma1}
r_{\rm L\gamma syn1,6}\sim1.7\times10^8\alpha^{-3/7}\xi^{6/7}L_{\rm th,28}^{-6/7}\nu_{\rm obs,keV}^{-3/7}T_{\rm pc,6}^{6/7}R_{\rm open,lc}^{1/7}P_0^{4/7}B_{\rm s,12}^{3/7},~~(t_{\rm ad}>t_{\rm cool})
\end{eqnarray}
and
\begin{eqnarray}\label{sec4:r_ngamma2}
r_{\rm L\gamma syn2,6}\sim2.3\alpha^{3/7}\xi^{-3/7}L_{\rm th,28}^{3/7}\nu_{\rm obs,keV}^{3/7}T_{\rm pc,6}^{-3/7}R_{\rm open,lc}^{-1/14}P_0^{-2/7}B_{\rm s,12}^{3/7}.~~(t_{\rm ad}<t_{\rm cool})
\end{eqnarray}

\begin{figure}
 \begin{center}
  \includegraphics[width=100mm, angle=270]{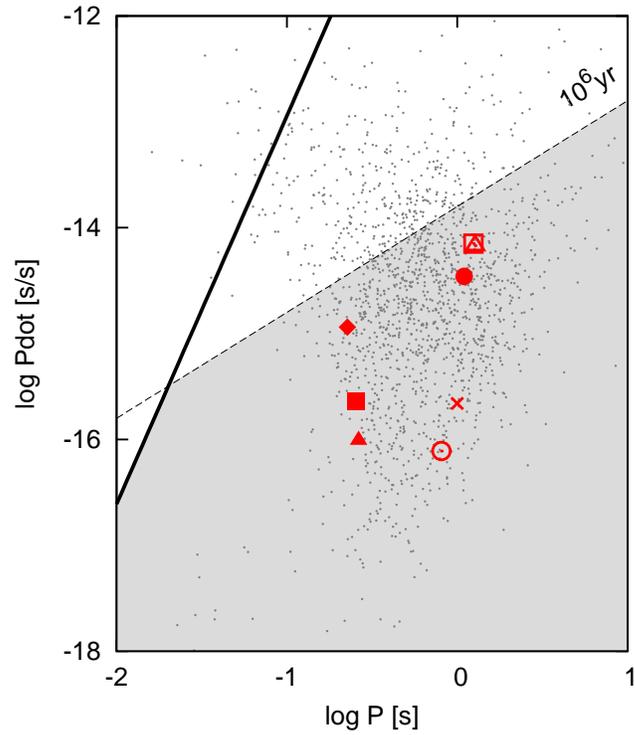}
 \end{center}
 \caption{Synchrotron radiation death line on the $P$-$\dot{P}$ diagram. Secondary particles are produced by the ingoing curvature photons via the photon-photon pair production. The solid line denotes the death line obtained from inequality (\ref{sec4:ingammagammaPdotP}). Each point is same as in figure \ref{fig1}. Dashed line is same as in figure \ref{fig3}. Our samples locate the filled region (below the death line) so that the model in section 4.2 does not explain their non-thermal X-ray emissions.}
 \label{fig8}
\end{figure}

From equations (\ref{sec4:r_gamma}), (\ref{sec4:r_ct}), (\ref{sec4:r_gammagamma}), (\ref{sec4:r_ngamma1}) and (\ref{sec4:r_ngamma2}), we obtain the relation $\max\{r_{\rm ct}, r_{\rm L\gamma syn1}\}<r_{\rm sec}<\min\{r_{\gamma{\rm syn}}, r_{\gamma\gamma}, r_{\rm L\gamma syn2}\}$. For the old pulsars, the relations $r_{\rm ct}<r_{\rm L\gamma syn1}$ and $r_{\rm L\gamma syn2}<r_{\gamma{\rm syn}}$ are satisfied. Therefore, we obtain the constraint for the synchrotron radiation region $r_{\rm sec}$ as
\begin{eqnarray}\label{ingammagamma}
r_{\rm L\gamma syn1}<r_{\rm sec}<\min\{r_{\gamma\gamma},r_{\rm L\gamma syn2}\}
\end{eqnarray} 
If we only consider the condition $r_{\rm L\gamma syn2}/r_{\rm L\gamma syn1}>1$ as a death line, this is given by
\begin{eqnarray}\label{sec4:ingammagammaPdotP}
\dot{P}>2.9\times10^{-10}\alpha^{-2/3}\xi\varepsilon_{\rm th,-3}^{-1}\nu_{\rm obs,keV}^{-2/3}T_{\rm pc,6}R_{\rm open,lc}^{1/6}P_0^{11/3}~{\rm s~s}^{-1},
\end{eqnarray}
where $\varepsilon_{\rm th}\equiv L_{\rm th}/L_{\rm sd}$. We show this condition with $\xi=\varepsilon_{\rm th,-3}=\nu_{\rm obs,keV}=T_{\rm pc,6}=R_{\rm open,lc}=1$ in figure \ref{fig8} as a solid line. The filled region shows the parameter region whose pulsars does not satisfy the condition $r_{\rm L\gamma syn1} < r_{\rm L\gamma syn2}$ so that the observed non-thermal X-ray emissions are not explained. Clearly, there are no old pulsars which can satisfy above condition. Note that the result, inequality (\ref{sec4:ingammagammaPdotP}), strongly depends on the thermal luminosity included in the ratio $\xi$ and the efficiency $\varepsilon_{\rm th}$, and these values are observationally constrained. Since we do not consider the thermal emission from the bulk surface of the neutron star, the obtained results are only valid for old pulsars with $\gtrsim1$Myr. We conclude that the ingoing case does not explain the observed luminosity of the non-thermal component, which exceeds or is the same value as the luminosity of thermal component.

\section{DISCUSSION}

In this paper, we investigate the synchrotron radiation from particles which are produced by ingoing or outgoing primary particles. For the particle production processes, we consider both the magnetic and the photon-photon pair productions.

Our model has a number of the optimistic assumptions to enlarge the luminosity of the synchrotron radiation. Our assumed Lorentz factor of primary particles $\gamma_{\rm p}$ is the maximum value in principle (equation \ref{sec2:gamma_p}). The Lorentz factor $\gamma_{\rm p}$ is supposed to be determined by the force balance between the electric field acceleration and the radiation reaction force. Then, the value of Lorentz factor $\gamma_{\rm p}$ is generally smaller than that of equation (\ref{sec2:gamma_p}). Our assumed value of Lorentz factor $\gamma_{\rm p}$ makes the upper limits for the emission locations $r_{\rm B\gamma}, r_{\rm \gamma syn}$, $r_{\rm \gamma spa}$, $r_{\rm LBsyn}$  and $r_{\gamma\gamma}$ (for the outgoing case) large and the lower limits for the emission regions $r_{\rm LBspa}$ and $r_{\gamma\gamma}$ (for the ingoing case) small. Other limits $r_{\rm cut}$, $r_{\rm ct}$, $r_{\rm L\gamma syn1}$ and $r_{\rm L\gamma syn2}$ do not depend on the Lorentz factor $\gamma_{\rm p}$ so that our obtained limits are the most optimistic values to explain the non-thermal X-ray emission by the synchrotron radiation.

We also assume the extreme case that the kinetic energy flux of primary particles is equal to the spin-down luminosity in the outgoing primary particle case. This means that a half of the Poynting flux is converted to the particle kinetic energy flux at the neutron star surface. Our adopted values of the optical depth for the magnetic pair production $\tau_{\rm B\gamma}\sim1$ and the cross section of the photon-photon collision $\sigma_{\gamma\gamma}\sim0.2\sigma_{\rm T}$ also correspond to the maximum values. We further assume the much effective pair cascade for the magnetic pair production case, i.e., the number of produced pairs increases by $\gamma_{\rm s,pair}/\gamma_{\rm s,lt}$ times (inequality \ref{sec2:number2}). These assumptions for the particle production make the upper limits $r_{\rm LBsyn}$ and $r_{\rm L\gamma syn2}$ large and the lower limits $r_{\rm LBspa}$ and $r_{\rm L\gamma syn2}$ small. Therefore, we argue that pulsars which locate beyond our obtained death lines on $P$-$\dot{P}$ diagram (the filled region) are difficult to explain the observed non-thermal component by the synchrotron radiation in the dipole magnetic field. 

Note that we only consider the constraints from energetics. Considering more detailed electrodynamics such as realistic current distribution flowing the magnetosphere should give more stringent limits for the non-thermal X-ray luminosity. This will be discussed in separate papers.

\subsection{Effects of Strong and Small-Scale Magnetic Field}

So far, we have assumed that all pulsars have only the dipole magnetic field component. Strong and small-scale magnetic field as compared to the global dipolar field seems to exist at the neutron star surface (e.g., Ruderman \& Sutherland 1975, Geppert et al. 2013). Here we briefly discuss the effects of such a small-scale magnetic field.

\begin{figure}
 \begin{center}
  \includegraphics[width=100mm, angle=270]{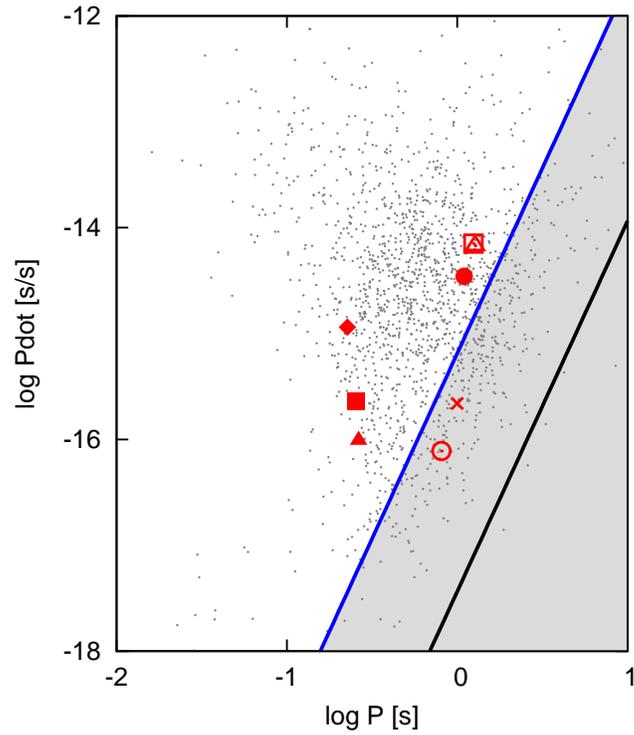}
 \end{center}
 \caption{Synchrotron radiation death line on the $P$-$\dot{P}$ diagram. Secondary particles are produced by the outgoing curvature photons via the magnetic pair production including the effects of strong and small-scale magnetic field. The black and blue lines denote the death lines obtained from inequalities (\ref{sec5:outBgammaPdotP}) and (\ref{sec5:outBgammaPdotP2}), respectively. Each point is same as figure \ref{fig1}. The oldest pulsar in our samples, J0108-1431, locates above the death line of inequality (\ref{sec5:outBgammaPdotP}) but below the death line of inequality (\ref{sec5:outBgammaPdotP2}) (the filled region) so that the model in section 5.1 does not explain their non-thermal X-ray emissions when we consider the effect of the radiation reaction force.} 
 \label{fig10}
\end{figure}

We consider the case of the outgoing primary particles with the magnetic pair production. We use the one-zone approximation for the locations of the curvature and the synchrotron radiations. Since a configuration of small-scale surface magnetic field shortens the curvature radius compared with the dipole one, the value of the pitch angle would be large at the inner region of the magnetosphere even if we consider the case of outgoing primary particles. Considering this effect, we take the curvature radius $R_{\rm cur}\sim r$ and the pitch angle $\alpha\sim 1$. Then, the same description for $r_{\rm ct}$ (equation \ref{sec4:r_ct}) also gives the lower limit for the emission location in this model. The constraint for the non-thermal X-ray luminosity gives the upper limit for the emission region as, 
\begin{eqnarray}\label{sec5:r_nB}
r_{\rm LBsyn,6}\sim39\alpha^{3/5}\varepsilon_{\rm syn,-3}^{-2/5}\nu_{\rm obs,keV}^{1/5}P_0^{-12/5}B_{\rm s,12}^{7/5}.
\end{eqnarray}
We find that this value is much larger than that of equation (\ref{sec3:r_nB}). On the other hand, the upper limit $r_{\rm B\gamma}$ is described as 
\begin{eqnarray}\label{sec5:r_Bgamma}
r_{\rm B\gamma}\sim8.8\alpha^{1/4}P_0^{-3/2}B_{\rm s,12}.
\end{eqnarray}
Then, we obtain the relation $\min\{r_{\rm LBsyn},r_{\rm B\gamma}\}=r_{\rm B\gamma}$ for our samples. The derived death line $r_{\rm B\gamma}/r_{\rm ct}>1$ is almost consistent with the pair production death line derived by \citet{RS75} as
\begin{eqnarray}\label{sec5:outBgammaPdotP}
\dot{P}>3.7\times10^{-18}\alpha^{-7/4}\nu_{\rm obs,keV}^{-1}P_0^{7/2}.
\end{eqnarray}
We depict the result with $\alpha=\nu_{\rm obs,keV}=1$ in figure \ref{fig10} as a black line. Most pulsars including our samples locate within the obtained death line in $P$-$\dot{P}$ diagram. 

However, the value of Lorentz factor $\gamma_{\rm p}$ should be determined by the local electric field structure and the balance of the radiation dragging force. Actual estimate decrease the value of Lorentz factor $\gamma_{\rm p}$, which is proportional to $R_{\rm cur}^{1/2}$. Since $R_{\rm cur}\sim r$ in our assumption, at the inner magnetosphere ($r\sim10^6$cm), the curvature radius is $\sim100$ times smaller than that of the dipole case [$R_{\rm cur}(r=R_{\rm NS})\sim10^8$cm]. Then, a small value of the Lorentz factor $\gamma_{\rm p}$ ($\sim0.1$ times) is expected when we use the same value of the accelerating electric field for both cases. If we take $\gamma_{\rm p}\rightarrow 0.1\gamma_{\rm p}$, the death line (inequality \ref{sec5:outBgammaPdotP}) is modified as
\begin{eqnarray}\label{sec5:outBgammaPdotP2}
\dot{P}>6.5\times10^{-16}\alpha^{-7/4}\nu_{\rm obs,keV}^{-1}P_0^{7/2}.
\end{eqnarray}
We depict this result as a blue line in figure \ref{fig10}. The oldest pulsar in our samples, J0108-1431, locates beyond this death line (the filled region in figure \ref{fig10}) on the $P$-$\dot{P}$ diagram. Therefore, even if we take the effect of the non-dipole component, the synchrotron radiation is difficult to explain the observed non-thermal X-ray emission from pulsars with the spin-down luminosity $L_{\rm sd,31}\lesssim(1-10)$.

\section{CONCLUSIONS}

We study the synchrotron radiation as the emission mechanism of the observed non-thermal X-ray component from old pulsars. We assume that the power-law component of the observed X-ray spectra is caused by the synchrotron radiation from produced electrons and positrons in the magnetosphere. We consider two pair production mechanisms of X-ray emitting particles, the magnetic and the photon-photon pair productions. High-energy photons, which ignite the pair production, are emitted via the curvature radiation of the primary particles. We use the analytical description for the radiative transfer and the pair production threshold to calculate the allowed range of the emission location. 

In the case of the outgoing primary particles and the magnetic pair production, for pulsars with the spin-down luminosity $L_{\rm sd,31}\lesssim(10-100)$, our model is difficult to explain their observed non-thermal emission based on the dipole magnetic field structure. Pulsars with the spin-down luminosity $L_{\rm sd,31}\gtrsim (10-100)$ are explained by this model. In our samples of old pulsars, the allowed range of the locations of the particle acceleration and the synchrotron radiation is $2\lesssim r_6\lesssim40$, which is closer to the neutron star surface than that of the outer gap. This emission region is almost similar to the radio emission region (e.g., Hassall et al. 2012). 

In the case of the outgoing primary particles and the photon-photon pair production, the pair-production optical depth for the curvature photons emitted from primary particles is too low to explain the effective number of secondary particles required from the luminosity of observed non-thermal component. The optical depth is mainly determined by the collision angle between a curvature photon and a thermal photon and the number density of the thermal photons at the particle-producing region. The collision angle is small near the neutron star surface. For the outer region of the magnetosphere, the number density of the thermal photons decreases so that the optical depth also decreases. Our model indicates that the non-thermal luminosity does not exceed the thermal one for old pulsars. 

If we consider the ingoing case and the magnetic pair production, the observed thermal luminosity from the polar cap strongly constraints the number flux of the ingoing primary particles. Then, the ingoing cases which were previously suggested by some authors (e.g., Zhang \& Cheng 1997, Wang et al. 1998) only explain the luminosity of non-thermal component for pulsars with the spin-down luminosity $L_{\rm sd,31}\gtrsim100$.

In the case of the ingoing primary particles and the photon-photon pair production, we assume the collision angle between a curvature photon and a thermal photon is $\cos\theta_{\rm col}\sim-1$. The observed thermal luminosity from the polar cap surface also strongly constraints the number flux of the primary particles. Then, the effective number of the secondary particles is limited to much small value. As a result, the non-thermal X-ray luminosity from the synchrotron radiation is much lower than the thermal ones. The synchrotron radiation from the secondary particles produced by the photon-photon pair production is difficult to explain the observed non-thermal component. 

In order to explain the observed non-thermal X-ray emission based on the synchrotron radiation, we conclude that the region of the particle acceleration should locate at the inner region of magnetosphere $r_6\lesssim10$ where the magnetic pair production can occurs for pulsars with the spin-down luminosity $L_{\rm sd,31} \sim(10-100)$. Note that the outer-gap death line is also locate at $L_{\rm sd,31}\sim(10-100)$ \citep{WH11}. Therefore, we suggest that when pulsars cross the outer-gap death line, their location of the particle acceleration move to inner magnetosphere where the magnetic pair production occurs. 

For the non-thermal emission from pulsars with the spin-down luminosity $L_{\rm sd,31}\lesssim10$ and the ratio $\xi\gtrsim1$ such as J0108-1431, other emission mechanisms such as the inverse Compton scattering \citep{ZH00} are required. Note that the inverse Compton model also requires that the particle acceleration occurs at inner region of the magnetosphere ($r_6\lesssim10$) because the model of \citet{ZH00} is based on the existence of the polar cap accelerator and the resonant scattering region for the X-ray is $r_6\sim10$. 

\section*{Acknowledgements}

We thank K. Asano, S. Shibata, J. Takata, Y. Teraki and T. Terasawa for valuable discussion and comments. We also thank our anonymous referee for helpful comments. This work is supported in part by Grant-in-Aid for Scientific Research on Innovative Areas (S.K., 24103006) and JSPS Research Fellowships for Young Scientists (S.J.T., 2510447). 


\clearpage
\begin{table}
\rotatebox{90}{
\begin{minipage}{\textheight}
\begin{center}
\begin{tabular}{cccccccc}
\multicolumn{8}{c}{TABLE 1 Pulsar observed parameters} \\ \hline
Name & J0108-1431 & B0628-28 &B0834+06 & B0943+10 & B0950+08 & B1451-68 & B1929+10 \\ \hline
$P$(s) & $0.808$ & $1.24$ & $1.27$ & $1.10$ & $0.253$  & $0.263$ & $0.227$ \\
$B_{\rm s}$($10^{12}$G) & $0.50$ & $6.0$ & $6.0$ & $4.0$ & $0.48$ & $0.32$ & $1.0$ \\
$L_{\rm syn}$($10^{28}$erg s$^{-1}$) & $4.1$ & $33$ & $7.8$ & $15$ & $83$ & $50$ & $120$ \\
$L_{\rm th}$($10^{28}$erg s$^{-1}$) & $3.7$ & $7.0$ & $7.8$ & $15$ & $16$ & $15$ & $42$ \\
$T_{\rm pc}$($10^6$K) & $1.3$ & $3.3$ & $1.6$ & $3.2$ & $1.8$ & $4.1$ & $3.5$ \\
Ref. & 1 & 2 & 3 & 4 & 5 & 6 & 7 \\ \hline
\multicolumn{8}{l}{}%
\label{tab:sample}
\end{tabular}
\end{center}
NOTES.- Period is taken from ATNF Pulsar Catalog \citep{Ma05}\footnote{http://www.atnf.csiro.au/research/pulsar/psrcat}. For the magnitude of surface magnetic field, we use period and its derivative taken from ATNF catalogue and calculate $\mu_{\rm mag}^2\equiv(B_{\rm s}R_{\rm NS}^3/2)^2=3Ic^3P\dot{P}/8\pi^2$. Observed values of non-thermal X-ray luminosity is taken from \citet{Po12b}. For the values of thermal luminosity and temperature, we take from following references; (1)\citet{Po12b} (2)\citet{TO05} (3)\citet{Gi08} (4)\citet{He13}
(5)\citet{ZP04} (6)\citet{Po12a} (7)\citet{MPG08}.
\end{minipage}
}
\end{table}


\begin{table}
\rotatebox{90}{
\begin{minipage}{\textheight}
\begin{center}
\begin{tabular}{llllllc}
\multicolumn{7}{c}{TABLE 2 Definitions of our used limits for locations of emission region.} \\ \hline
Location & Constraint & Limit & Pair production & Frequency & Timescale & Equation \\ \hline
$r_{\rm \gamma syn}$ & Energy of secondary particles & Upper & Not depend & $\nu_{\rm obs}>\nu_{\rm ct}$ & Not depend & (\ref{sec2:energy}) \\
$r_{\rm \gamma spa}$ & Energy of secondary particles & Upper & Not depend & $\nu_{\rm obs}<\nu_{\rm ct}$ & Not depend & (\ref{sec2:energy}) \\
$r_{\rm ct}$ & Characteristic frequency & Lower & Not depend & $\nu_{\rm obs}>\nu_{\rm ct}$ & Not depend & (\ref{sec2:frequency}) \\
$r_{\rm cut}$ & Characteristic frequency & Lower & Not depend & $\nu_{\rm obs}<\nu_{\rm ct}$ & Not depend & (\ref{sec2:frequency2}) \\
$r_{\rm B\gamma}$ & Production threshold & Lower & Magnetic & Not depend & Not depend & (\ref{sec2:Bcreation}) \\ 
$r_{\gamma\gamma}$ & Production threshold & Lower or upper & Photon-photon & Not depend & Not depend & (\ref{sec2:gcreation}) \\ 
$r_{\rm LBsyn}$ & Non-thermal luminosity & Upper & Magnetic & $\nu_{\rm obs}>\nu_{\rm ct}$ & $t_{\rm cool}<t_{\rm ad}$ & (\ref{sec2:number2}) \\ 
$r_{\rm LBspa}$ & Non-thermal luminosity & Lower & Magnetic & $\nu_{\rm obs}<\nu_{\rm ct}$ & $t_{\rm cool}<t_{\rm ad}$ & (\ref{sec2:number2}) \\ 
$r_{\rm L\gamma syn1}$ & Non-thermal luminosity & Lower & Photon-photon & $\nu_{\rm obs}>\nu_{\rm ct}$ & $t_{\rm cool}<t_{\rm ad}$ & (\ref{sec2:number}) \\ 
$r_{\rm L\gamma syn2}$ & Non-thermal luminosity & Upper & Photon-photon & $\nu_{\rm obs}>\nu_{\rm ct}$ & $t_{\rm cool}>t_{\rm ad}$ & (\ref{sec2:number}) \\  
$r_{\rm L\gamma spa1}$ & Non-thermal luminosity & Lower & Photon-photon & $\nu_{\rm obs}<\nu_{\rm ct}$ & $t_{\rm cool}<t_{\rm ad}$ & (\ref{sec2:number}) \\  \hline
\multicolumn{7}{l}{}%
\label{tab:gamma2}
\end{tabular}
\end{center}
\end{minipage}
}
\end{table}


\clearpage

\clearpage

\clearpage

\clearpage

\clearpage

\clearpage

\clearpage
\appendix

\section{EMISSION LOCATION OF SECONDARY PARTICLES}

We calculate the relation between two distances to the emission locations of the ingoing primary and the secondary particles. We assume that the structure of magnetic field is static dipole. We ignore the effects of aberration and light bending due to gravity.

\begin{figure}
 \begin{center}
  \includegraphics[width=100mm]{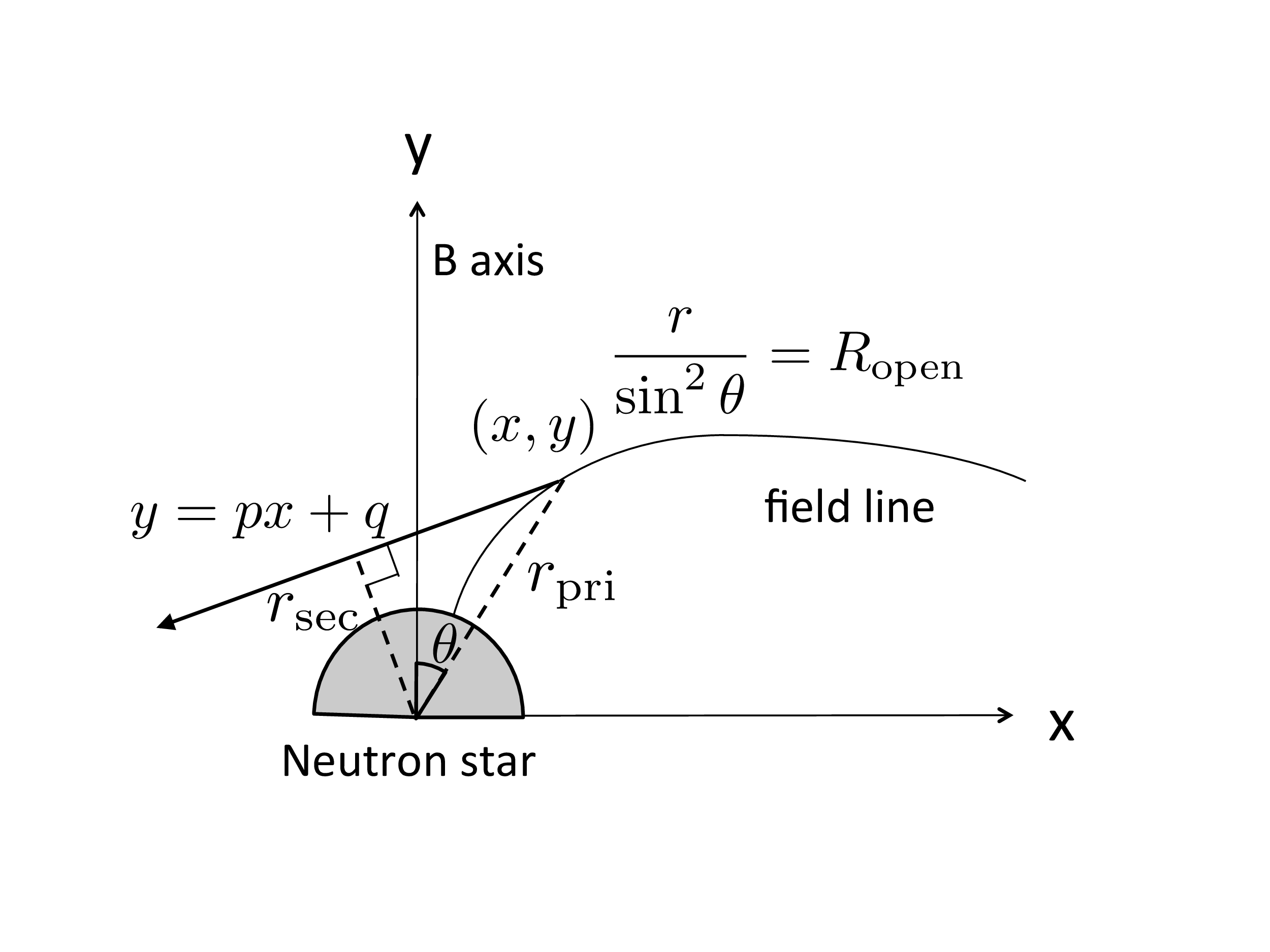}
 \end{center}
 \caption{Schematic picture on the meridional plane. The magnetic axis corresponds to $y$-axis. A curved line denote a magnetic field line and a Cartesian coordinate ($x,y$) is the emission point of the curvature radiation. Thick arrow denotes the trajectory of a curvature photon, which is a tangential line to the magnetic field line. For the emission region of secondary particles $r_{\rm sec}$, we assume the point whose distance from the centre of the neutron star is minimum. } 
 \label{figA1}
\end{figure}

Schematic picture is shown in figure \ref{figA1}. We consider that a particle moves along a given magnetic field line. The magnetic axis corresponds to the $y$-axis and a primary particle locates at a point $(x,y)$ in figure \ref{figA1}. A field line (a curved line in figure \ref{figA1}) is characterized in the following equation, 
\begin{eqnarray}\label{app:R_open}
\frac{r}{\sin^2\theta}=R_{\rm open}\ (\ge R_{\rm lc}),
\end{eqnarray}
where $r=\sqrt{x^2+y^2}$, $\theta$ denotes the co-latitude of the point from the magnetic axis and $R_{\rm open}$ is maximum distance to a point on a magnetic loop. We only consider within the open zone in the magnetosphere so that $R_{\rm open}$ is always larger than the radius of the light cylinder $R_{\rm lc}$. We assume that the distance $r_{\rm sec}$ is the minimum distance from the centre of the neutron star to a point on the trajectory of the curvature photons emitted by a primary particle $r_{\rm pri}$ on a magnetic field line with $R_{\rm open}$. As long as the distance $r_{\rm sec}$ is larger than the radius of the star, we can detect the photons emitted from secondary particles in principle. Here, we introduce the normalized distance $r_{\rm pri,op}\equiv r_{\rm pri}/R_{\rm open}$. The point in Cartesian coordinate for a primary particle is described by 
\begin{eqnarray}\label{app:xy}
(x,y)=\left(r_{\rm pri}\sqrt{r_{\rm pri,op}},r_{\rm pri}\sqrt{1-r_{\rm pri,op}}\right).
\end{eqnarray}
The unit vector of the direction of the magnetic field at a point $(x,y)$ is written as
\begin{eqnarray}\label{app:bxby}
{\bf b}=(b_x,b_y)=\left(\frac{3\sqrt{r_{\rm pri,op}(1-r_{\rm pri,op})}}{\sqrt{4-3r_{\rm pri,op}}},\frac{2-3r_{\rm pri,op}}{\sqrt{4-3r_{\rm pri,op}}}\right).
\end{eqnarray}
We calculate the coefficients of the equation of the trajectory $y=px+q$ of the photons emitted at a point $(x,y)$ as
\begin{eqnarray}\label{app:p}
p=\frac{2-3r_{\rm pri,op}}{3\sqrt{r_{\rm pri,op}(1-r_{\rm pri,op})}},
\end{eqnarray}
\begin{eqnarray}\label{app:q}
q=\frac{r_{\rm pri}}{3\sqrt{1-r_{\rm pri,op}}}.
\end{eqnarray}
Thus, we derive the location of the emission region of the secondary particles as 
\begin{eqnarray}\label{app:r_sec}
r_{\rm sec}=\sqrt{\frac{q^2}{1+p^2}}.
\end{eqnarray}
Using equations (\ref{app:p}), (\ref{app:q}) and (\ref{app:r_sec}), we describe the relation between the locations $r_{\rm pri}$ and $r_{\rm sec}$ as,
\begin{eqnarray}\label{app:r_pri}
r_{\rm pri}^3+3r_{\rm pri}r_{\rm sec}^2-4R_{\rm open}r_{\rm sec}^2=0.
\end{eqnarray}
Note that the second term in the left-hand side of equation (\ref{app:r_pri}) is neglected as long as $\sqrt{3}r_{\rm sec}<r_{\rm pri}<(3/4)R_{\rm open}$. 
The upper limit $(3/4)R_{\rm open}$ is nearly or larger than the radius of the light cylinder so that the condition $r_{\rm pri}<(3/4)R_{\rm open}$ is always satisfied in our model. For the lower limit, even if we take the distance $r_{\rm sec}=R_{\rm NS}$, the ratios are $r_{\rm pri,6}\sim12$ for $P_0=0.1$ and $r_{\rm pri,6}\sim$27 for $P_0=$1. Then, the condition $r_{\rm pri}>\sqrt{3}r_{\rm sec}$ is always satisfied for old pulsars. Therefore, we approximately describe $r_{\rm pri}$ as
\begin{eqnarray}\label{app:approxr_pri}
r_{\rm pri,6}>27r_{\rm sec,6}^{2/3}R_{\rm open,lc}^{1/3}P_0^{1/3}.
\end{eqnarray}


\clearpage

\end{document}